\documentclass[sigconf,natbib=false,nonacm]{acmart}
\usepackage[ruled,vlined]{algorithm2e}
\usepackage{multirow}
\usepackage{algpseudocode}
\usepackage{tabularx}
\usepackage{enumitem}
\usepackage{changepage}
\usepackage{appendix}
\setlist[itemize]{leftmargin=*}
\newcolumntype{Y}{>{\centering\arraybackslash}X}

\AtBeginDocument{%
  }

\RequirePackage[
  datamodel=acmdatamodel,
  style=acmnumeric,
  ]{biblatex}

\begin{document}
\title{Unraveling the MEV Enigma: ABI-Free Detection Model using Graph Neural Networks}

\author{Seongwan Park}
\authornote{Both authors contributed equally to this research.}
\affiliation{%
  \institution{Seoul National University}
  \country{South Korea}
}
\email{sucre87@snu.ac.kr}

\author{Woojin Jeong}
\authornotemark[1]
\affiliation{%
  \institution{Seoul National University}
  \country{South Korea}
}
\email{jwj7955@snu.ac.kr}

\author{Yunyoung Lee}
\affiliation{%
  \institution{Seoul National University}
  \country{South Korea}
}
\email{tommja@snu.ac.kr}

\author{Bumho Son}
\affiliation{%
  \institution{Seoul National University}
  \country{South Korea}
}
\email{andymogul@snu.ac.kr}

\author{Huisu Jang}
\affiliation{%
  \institution{Soongsil University}
  \country{South Korea}
}
\email{yej523@ssu.ac.kr}

\author{Jaewook Lee}
\affiliation{%
  \institution{Seoul National University}
  \country{South Korea}
}
\email{jaewook@gmail.com}


\begin{abstract}

The detection of Maximal Extractable Value (MEV) in blockchain is crucial for enhancing blockchain security, as it enables the evaluation of potential consensus layer risks, the effectiveness of anti-centralization solutions, and the assessment of user exploitation. However, existing MEV detection methods face limitations due to their low recall rate, reliance on pre-registered Application Binary Interfaces (ABIs) and the need for continuous monitoring of new DeFi services. 

In this paper, we propose ArbiNet, a novel GNN-based detection model that offers a low-overhead and accurate solution for MEV detection without requiring knowledge of smart contract code or ABIs. We collected an extensive MEV dataset, surpassing currently available public datasets, to train ArbiNet. Our implemented model and open dataset enhance the understanding of the MEV landscape, serving as a foundation for MEV quantification and improved blockchain security.

\end{abstract}

\ccsdesc[500]{Security and privacy~Intrusion/anomaly detection and malware mitigation}
\ccsdesc[500]{Security and privacy~Distributed systems security}

\keywords{Maximal Extractable Value, Blockchain, Graph Neural Network}

\maketitle

\section{Introduction}

Consensus security and decentralization are essential aspects of public blockchains, such as Bitcoin and Ethereum. Since their inception, numerous studies have attempted to improve consensus security and promote decentralization. Past works have primarily focused on consensus mechanism\cite{buterin2017casper,gervais2016security, buchman2016tendermint}, mining pool strategy\cite{wang2019pool, yang2022if, azimy2022preventing, grunspan2020selfish,xiao2020modeling}, and incentive compatiblility of token economy\cite{schwarz2022three, chitra2022improving, lyu2022empirical, carlsten2016instability}.

Recently, MEV (Maximal Extractable Value) has emerged as another significant factor influencing consensus security and decentralization in blockchain networks \cite{barczentewicz2023mev, berg2022empirical}. The rapid growth of Decentralized Finance (DeFi) applications in Turing-complete blockchains, like Ethereum, has led to a substantial increase in the amount of MEV\cite{daian2019flash}. Moreover, since Ethereum's transition from Proof of Work (PoW) to Proof of Stake (PoS)\cite{mainnet_merge_announcement}, the block reward has been reduced by more than 90\%\cite{kapengut2023event}, which has amplified the influence of MEV on network security. This change might increase the risks of incentivizing miners to censor transactions in order to acquire more MEV.\cite{zhou2021just}

Toxic MEV transactions, such as sandwich attacks\cite{zhou2021high, eskandari2020sok, wang2022impact}, lead to unexpected user losses\cite{torres2021frontrunner,zhou2021a2mm}, and the amount of MEV in a block impacts miners' incentives to fork the chain. Additionally, the growing extractable value in a block has made efficient block building more sophisticated, necessitating the use of advanced hardware and strategies. This has resulted in the issue of MEV centralization, where skilled block builders earn more MEV profits than less sophisticated builders \cite{jensen2023multi}. In order to address the centralization of MEV and promote a more equitable distribution of profits, several mitigation strategies have been proposed, such as Flashbots' MEV-boost services \cite{MEV-Boost} and Ethereum's in-protocol Proposer-Builder Separation (PBS) \cite{PBS}.

Given the challenges associated with MEV, accurately measuring its impact on the ecosystem is crucial for enhancing blockchain security. Quantifying MEV helps assess the current state of MEV, evaluate potential security risks to the consensus layer, understand the actual impact of solutions against MEV centralization, and provide tools for modeling scenarios. However, the complex and diverse nature of MEV transactions makes it arduous to provide an explicit definition of MEV\cite{judmayerestimating, yang2022sok}, which in turn results in difficulties of its detection. Although theoretical definitions of MEV have been suggested\cite{obadia2021unity, mazorra2022price}, they are not practical for use in MEV detection.

Despite the inherent difficulties, several attempts have been made to accurately measure extractable value in blockchains. \cite{qin2022quantifying} made the first attempt using heuristics to identify various MEV transactions, while \cite{piet2022extracting} employed a unique graph algorithm to capture diverse forms of MEV transactions. 
\cite{weintraub2022flash} also developed measurement methodologies and shared the code. Although the previous studies had their unique contributions, they do have certain limitations. First, most of heuristics are overly conservative, leading to the underestimation of MEV and an incomplete understanding of its scope. Second, detecting MEV transactions typically requires knowledge of smart contracts, complicating maintenance. 

In this paper, we focus on sandwich and arbitrage transactions, which account for 99\% of total MEV \cite{flashbots_explorer}. We identify all events and function calls employed by MEV transactions, and classify these transactions based on their patterns and strategies. By applying this classification, we label MEV transactions and demonstrate that our collected data provides a more accurate representation of past MEV data than existing public sources. However, we also acknowledge that this approach still has same limitations with existing detection algorithms, regarding the knowledge of contract ABI. To address these limitations, we propose ArbiNet, a novel arbitrage detection model based on Graph Neural Network (GNN) layers, and a simple algorithm for detecting sandwiches. Both ArbiNet and the sandwich detection algorithm overcome the limitations of existing methods, offering future MEV detection solutions without requiring knowledge of smart contract ABIs or events. We emphasize that our ABI-free approaches not only contribute to efficient MEV quantification but also enable the rapid identification of emerging threats.

To summarize, our key contributions are:

\begin{itemize}[noitemsep,topsep=0pt]
\item We clearly identify the limitations of existing MEV detection algorithms, which have been rarely discussed.

\item We thoroughly examine and classify MEV transactions, including two categories of sandwich and five categories of arbitrage, effectively identifying novel strategies.

\item Building upon the forms identified in the previous point, we make a publicly available extensive dataset of MEV transactions, surpassing existing public MEV data. The dataset is also used for training and evaluating ArbiNet.

\item We implement ArbiNet, a low-overhead, high-performance arbitrage detection model that eliminates the need for event or ABI knowledge. It demonstrates significantly better detection performance (F1) compared to public data.
\end{itemize}

The rest of the paper is structured as follows. Section \ref{Background} provides essential background knowledge on MEV detection and GNN. Section \ref{Existing algorithm} introduces widely-used existing algorithms and thoroughly outlines their limitations. In Section \ref{Proposed Method}, we present our proposed models for MEV detection. Section \ref{classification} offers an in-depth exploration of the MEV transactions used for model training and evaluation. In Section \ref{evaluation}, we compare our collected MEV data with other sources and evaluate ArbiNet against the labeled MEV data. We discuss our model's limitations in Section \ref{discussion}, and conclude in Section \ref{Conclusion}.

\section{Background}\label{Background}
\subsection{Smart Contract \& ABI}
Smart contracts are programs that are deployed on Turing-complete blockchains, such as Ethereum\cite{wood2014ethereum}. Transactions can call functions implemented in smart contracts, and some function calls emit events. For example, if user A wants to send DAI, the user should call the  \emph{transfer(address recipient, uint256 amount)} function in the DAI contract, after which the \emph{event Transfer} is emitted. 

Decoding a transaction to investigate the called functions and emitted events requires the smart contract's ABI (Application Binary Interface) and the transaction receipt. The ABI specifies the methods that can be called on a smart contract, as well as the data types and encoding used for the method arguments and return values. Since the bytecode of the smart contract code is deployed instead of the actual contract code, the contract ABI is not directly accessible from the blockchain node. Although ABIs for standard token contracts like ERC-20 are readily available, accessing the ABI for a general smart contract typically relies on centralized API services like Etherscan. The contract ABI becomes accessible only if the contract deployer has agreed to make it public.

\subsection{Decentralized Finance}

\noindent \textbf{DEX.}
Decentralized exchanges (DEXs) enable users to trade assets without requiring a trusted intermediary, as discussed by \cite{xu2023sok}. DEXs are typically implemented with the automatic market maker (AMM) to facilitate trades and determine prices automatically. The spot price is commonly defined as the ratio of two token reserves in the liquidity pool. Liquidity providers deposit assets into the pool contract of a DEX and receive shares that represent their ownership of the pool. They earn fees collected in the pool proportional to their share of the total pool liquidity. When a user sends a swap transaction to the pool contract, the AMM calculates the amount of assets that the user will receive in exchange. The decentralized nature of the network can lead to price differences between liquidity pools. Arbitrageurs can take advantage of these price differences by buying a cryptocurrency on one DEX where it is relatively cheaper and then selling it on another DEX where it is relatively more expensive\cite{sylvester2022modeling}. It allows them to profit from the price differences between the two exchanges.

\begin{figure}[h]
  \centering
  \includegraphics[width=\linewidth]{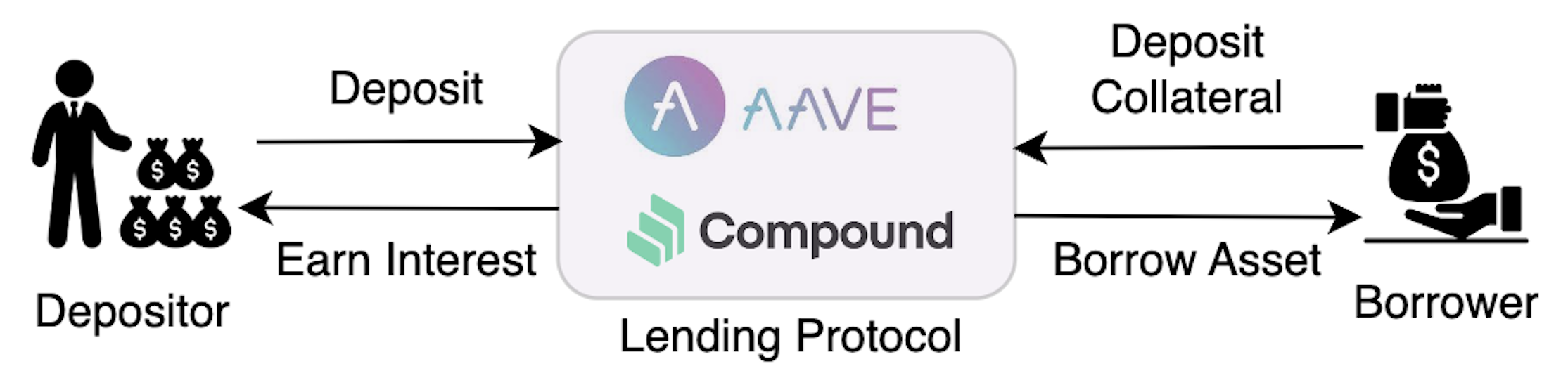}
  \caption{Lending protocol}
  \label{fig:Lending protocol}
\end{figure}

\noindent \textbf{Lending Protocol.}
A lending protocol manages a lending pool, which holds multiple lending and borrowing positions for various assets. The main entities involved in lending protocols are lenders, borrowers, and liquidators. Lenders deposit their capital into the lending pool and earn interest-bearing tokens as proof of deposit (Figure \ref{fig:Lending protocol}). For example, lender A deposits ETH and get cETH back from Compound pool. cETH serves as a certificate for A's original deposited asset and accrued interest. The lending pool stores all deposits, accessible to borrowers who provide collateral and borrow the desired assets. Borrowers can close their position any time once they repay the outstanding loan and interest, which unlocks the collateral.

\subsection{MEV}
\noindent \textbf{Arbitrage.}
Arbitrage is the most commonly used strategy for extracting MEV. Arbitrageurs exploit the price differences or market inefficiencies across different DEXs or other DeFi protocols. For example, if the price of ETH on Uniswap\cite{uniswap_whitepaper} is \$2,000 and the price of ETH on Sushiswap is \$2,100, an arbitrageur could buy ETH on Uniswap and immediately sell it on Sushiswap to make a profit of \$100 valued ETH. A real arbitrage example is described in Figure \ref{fig:Arbitrage1}. Three or more assets can also be involved in arbitrage transaction by trading across three different DEXs. As the complexity and variability of arbitrage strategies are increasing rapidly, defining the general strategies of arbitrage in MEV is obscure and accurately detecting the arbitrage transactions is challenging and time-consuming. We will describe the specific types of arbitrage transactions in further detail.

\noindent \textbf{Sandwich attack.   }
Sandwich attacks are a type of MEV transaction that are considered harmful, as they result in losses for users who intend to swap one token \emph{A} for another token \emph{B}. A sandwich attack typically consists of a front-run transaction and a back-run transaction executed by the attacker.

\begin{figure}[h]
  \centering
  \includegraphics[width=0.8\linewidth]{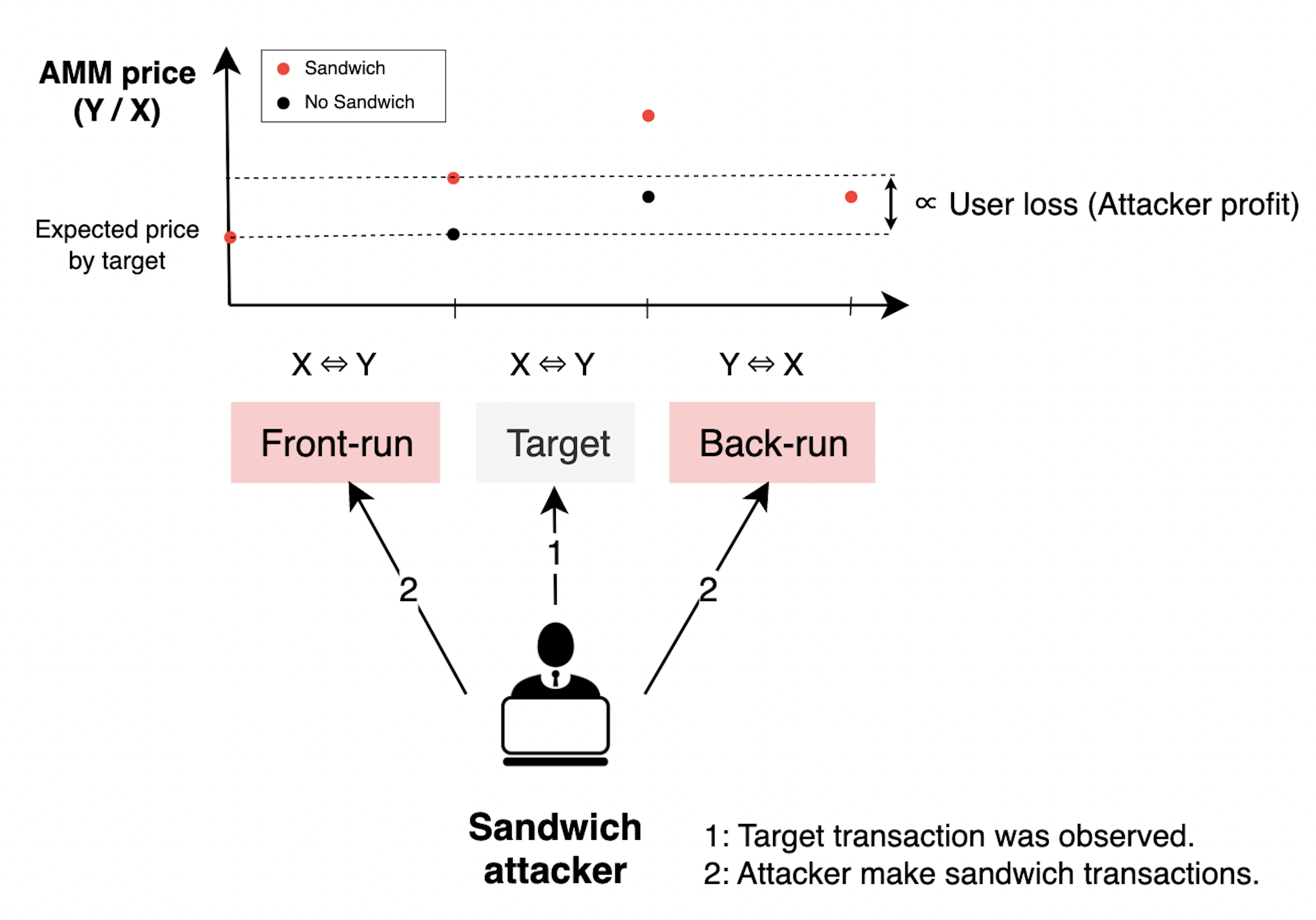}
  \caption{Sandwich attack}
  \label{fig:sandwich_back}  
\end{figure}

In a front-run transaction, the attacker preemptively swaps the same tokens, \emph{A} to \emph{B}  as the victim user intended. By doing so, the attacker's transaction is processed before the victim's transaction, causing a temporary increase in the token exchange ratio (\emph{A} / \emph{B}). As a result, the victim's transaction swaps tokens at a more expensive rate than initially anticipated, as shown in Figure \ref{fig:sandwich_back}.

Following the victim's transaction, the attacker executes a back-run transaction, selling token \emph{B} acquired from the front-run transaction. The attacker profits from the difference between the amount the victim initially intended to pay and the amount they ultimately paid.

\begin{figure}[h]
  \centering
  \includegraphics[width=0.7\linewidth]{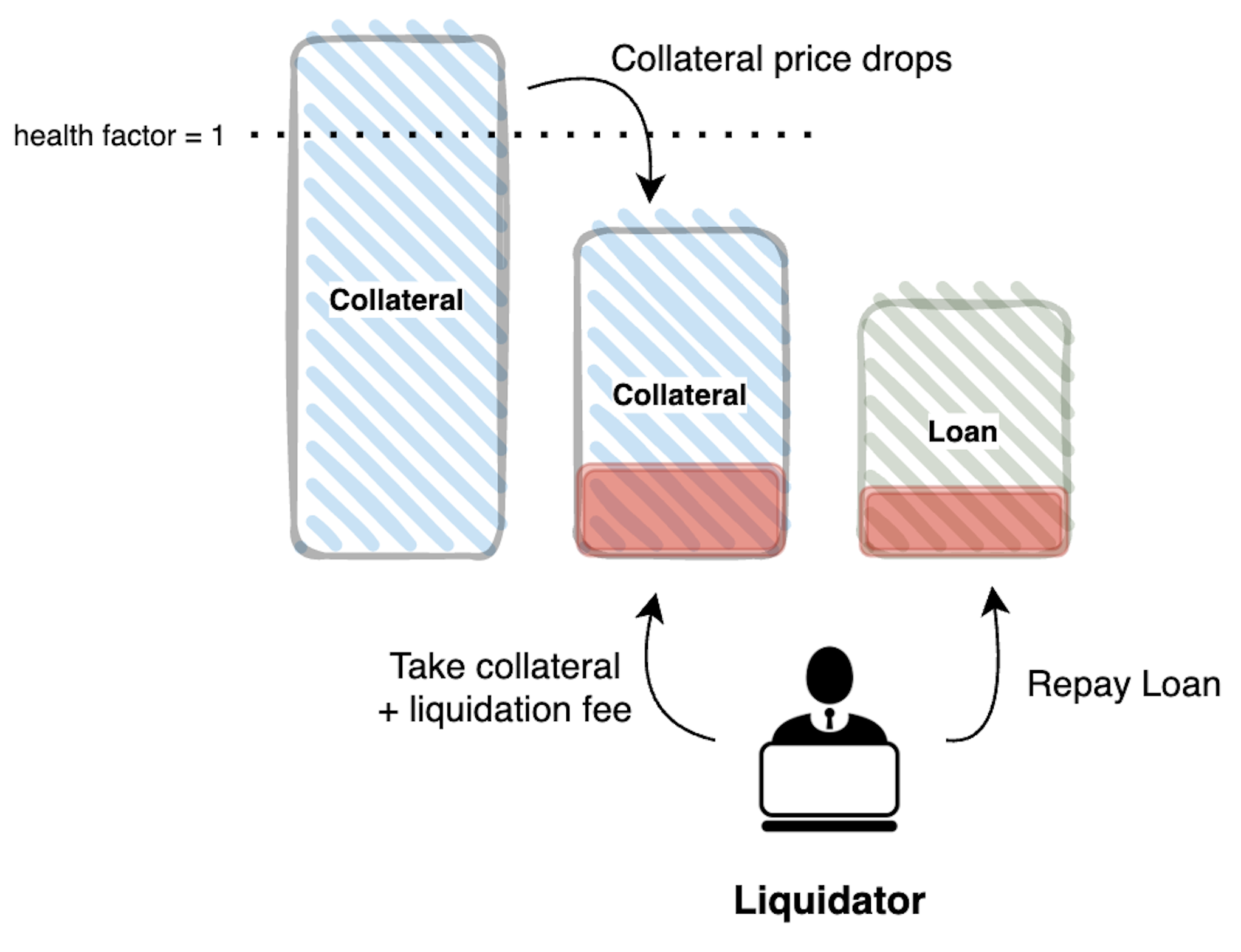}
  \caption{Liquidation Process}
  \Description{Liquidation Process}
  \label{fig:Liquidation Process}
\end{figure}

\noindent \textbf{Liquidation.   } 
Liquidation, another type of MEV, refers to the process of selling a borrower's collateral to repay their outstanding loan in the event of default. If the value of the borrower's deposited collateral falls below a certain threshold, liquidators can simultaneously pay the part of borrower’s loan and claim the part of borrower's collateral at a discounted price. This mechanism, presented in Figure \ref{fig:Liquidation Process} is critical for maintaining the stability of lending pools, where borrowers can deposit their assets as collateral to borrow other assets from the pool. Liquidation also provides an incentive for liquidators to promptly identify and liquidate unhealthy positions within the pool, thereby mitigating potential risks to the pool's stability.

\subsection{Graph Neural Network}
Graph neural networks (GNNs) are neural network-based models for graph representation that capture the structural information within graph data. Each node feature is sequentially updated by passing through hidden layers aggregating the neighbor nodes features in previous layer. The general learning process of GNN models can be represented by the following equation, where $h_v^k$ is the embedding of node $v$ in the $k^{th}$ layer, $\lvert N(v) \rvert$ is the number of neighbors of node $v$, $\sigma$ is the activation function (e.g., ReLU, sigmoid), and the learnable parameters are denoted as $W_k$ and $B_k$. 
$$ h_v^k = \sigma(W_k \sum_{u \in N(v)} {h_u^{k-1} \over \lvert N(v) \rvert}+B_kh_v^{k-1})$$
The parameter learning process in graph neural networks is similar to that of other neural network architectures, with learnable parameters adjusted through a process of iterative optimization to minimize the loss function. 

Graph Convolutional Network (GCN) \cite{kipf2016semi}, GraphSAGE \cite{hamilton2017inductive}, and Graph Attention Network (GAT)\cite{velivckovic2017graph} are three widely used GNN models for graph classification tasks, leveraged for detecting MEV transactions in this paper. These models differ in how node embeddings are aggregated, but all assume the homogeneous graph inputs.
GCN uses shared weight parameters to update the features of all nodes in a graph, which improves the model's performance and computing efficiency. GraphSAGE concatenates the embeddings of a node with those of its neighbors, rather than summing them up. GraphSAGE has the inductive property of generating node embeddings for evolving graphs, which makes it a highly flexible and scalable model in the domain of large graphs. Graph Attention Network (GAT) introduces a self-attention mechanism inspired by sequence models, allowing the model to associate different weights with each neighbor and increase its capacity for modeling complex relationships between nodes. The selection of a GNN model in a study depends on the properties of the input graph and the analysis objectives, and it is usually determined by an empirical evaluation of model performance.


\section{Existing MEV detection algorithm and challenges}\label{Existing algorithm}

\subsection{Existing algorithms}
There have been several efforts to detect MEV transactions and quantify them. \cite{qin2022quantifying} classifies MEV transactions into sandwich, arbitrage, liquidations and quantify them for the first time with heuristics they developed. \cite{piet2022extracting} regards sandwich as another type of arbitrage and detect arbitrage with its own method. \cite{weintraub2022flash} follows the style of \cite{qin2022quantifying} for MEV transaction detection, and provide a simple conceptual definition of each type of MEV transactions. \cite{weintraub2022flash} and \cite{flashbots-mevispectpy} provide open-source MEV detection code, which requires an Ethereum archive node. 

Detection algorithms presented in \cite{qin2022quantifying}, \cite{piet2022extracting}, \cite{flashbots-mevispectpy}, \cite{weintraub2022flash} commonly consists of two steps : preprocessing step and detection step. In the preprocessing step, the proposed algorithms transform each transaction data into a token transfer graph. The token graph is then classified to MEV and non-MEV in the detection step.

\begin{figure}[h]
  \centering
  \includegraphics[width=\linewidth]{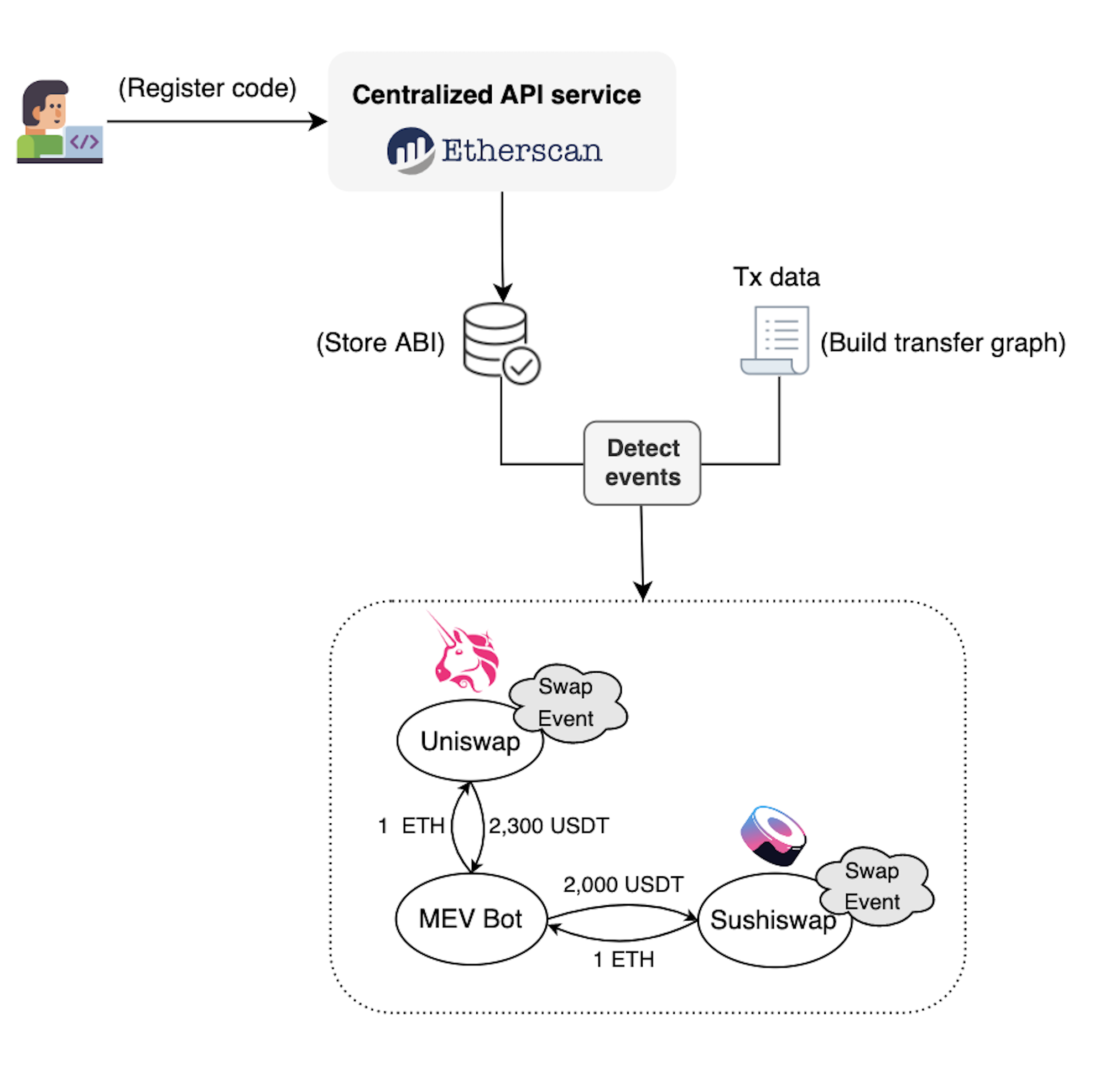}
  \caption{Preprocessing}
  \Description{The preprocessing of a transaction in today’s detection algorithm}
  \label{fig:Preprocessing}
\end{figure}

Figure \ref{fig:Preprocessing} provides a comprehensive overview of the preprocessing procedure in \cite{qin2022quantifying,piet2022extracting, flashbots-mevispectpy, weintraub2022flash}. In order to successfully preprocess a transaction to a a graph data, all the ABI (Application Binary Interface) or topic ids of relevant events of DeFi smart contracts are necessary. Since Ethereum nodes store bytecode of smart contracts instead of the original contract code, nodes do not have access to contract ABI. We note that decompiling bytecode to original code is unavailable, and centralized API services (\textit{e.g.} Etherscan\cite{etherscan}) are necessary in order to get ABI data. DeFi smart contract developers voluntarily register smart contract code to Etherscan, Etherscan verifies the raw code and provides ABI to users.

Each transaction data is decoded with stored ABI of DeFi contracts, and returns emitted events and called functions when executing the transaction. The transfer data of tokens also can be acquired using ERC-20 ABI\cite{erc20}, which is commonly used by most of fungible tokens issued on Ethereum.

With token transfer graph data, heuristics are used to detect each type of MEV transaction : sandwich, arbitrage, and liquidations.

\noindent \textbf{Sandwich. } 
\cite{qin2022quantifying} detects sandwich transactions $T_{A1}$ , $T_{A2}$ and victim transaction $T_V$  in the order $T_{A1}$, $T_V$, $T_{A2}$ using these heuristics: 1) $T_{A1}$ and $T_V$ transact $X \rightarrow Y$ and $T_{A2}$ transact $Y \rightarrow X$. 2) The amount of Y sold in $T_{A2}$ is in range of 90\% to 110\% of the amount of Y bought in $T_{A1}$. To detect token exchanges, they use swap events of 8 DEXs. \cite{weintraub2022flash} also use similar heuristics to \cite{qin2022quantifying}, with swap events from 5 DEXs. 

\noindent \textbf{Arbitrage. }
Arbitrage transactions have diverse forms, thus leading to higher detection error rate compared to sandwiches and liquidations. 

There have been two approaches to detect arbitrage transactions: one approach introduced by \cite{qin2022quantifying} is using only swap events, while the other approach introduced by \cite{piet2022extracting} is using token transfer graphs as well as swap events.

\cite{qin2022quantifying} uses heuristics that find loops between swap events from 8 DEXs. Then, heuristic checks if each loop satisfies following conditions : 1) output token of a swap event should be same to the input token of the next swap , 2) output amount of a swap event should be more than the input amount of the next swap. \cite{weintraub2022flash} implements this approach, which is publicly opened. One limitation of this approach is that, in order to get the exact amount and token of input and output, it requires investigating all the pools existing in DEXs. According to Coingecko\cite{coingecko}, there are 2405 distinct pools in Uniswap V2, and each pool has its distinct Ethereum address. Since it is unrealistic to store all pool addresses in all DEXs, it is necessary to request node twice additionally for each swap, which is inefficient.

On the other hand, algorithm in \cite{piet2022extracting} identifies cycles in token transfers and checks if each edge in a cycle is part of a currency exchange. It distinguishes intended currency exchanges such as swaps from a simple token transfer, by examining the emitted events. If two adjacent edges with different tokens are not part of a intended token exchange, the transaction is not considered as arbitrage. For example, a cycle of transfers in a transaction might be transfers of (1) A to B, (2) B to C, and (3) C to A. If there is a swap event that triggers (1) and (2), but no swap event accounts for (2) and (3), this transaction is not considered as arbitrage. 

\noindent \textbf{Liquidations. }
When liquidation occurs in a lending protocol like Aave and Compound, liquidation events are called. Unlike sandwiches and arbitrages, detecting liquidations only require tracking of liquidation function calls, without having to inspect token transfer graphs. \cite{weintraub2022flash} and \cite{flashbots-mevispectpy} keep track of liquidation events from 3 lending protocols: Aave V1, V2, and Compound.

We note that liquidations are not discussed for the rest of this paper for two reasons: 1) It is already easy to detect liquidations. We only need to keep track of a single liquidation function call, instead of inspecting on token transfer graphs and all the emitted events in a transaction.  2) Liquidations occupy a small portion of total MEV transactions, which are reported as  1\%  in Flashbots\cite{flashbots_explorer}.

\subsection{Challenges}

The above algorithms contain three significant limitations : dependency, maintenance, and recall rate.

\noindent \textbf{Dependency}
The current algorithm heavily depends on smart contract developers and centralized API services such as Etherscan. For the algorithm to be effective, contract developers must have registered their contract code with Etherscan beforehand. If not registered, there is no alternative method to decompile the bytecode and obtain the ABI. Moreover, if Etherscan does not provide the contract ABI to users, gaining access to ABI becomes difficult. This reliance on third-party services and the necessity of prior contract registration pose significant challenges to the algorithm's effectiveness.

\noindent \textbf{Maintenance}
As of April 2023, DeFiLlama \cite{defillama} reports at least 688 DeFi DApps on the Ethereum network, with numerous new DeFi services continuously emerging. Keeping up with all new DeFi services and updating the preprocessing logic accordingly can be challenging, as the current approach requires tracking every contract ABI used for MEV extraction. This method inevitably leads to frequent system fixes, with each update increasing code complexity and processing time.

\noindent \textbf{Recall rate}
The algorithms used in previous papers tend to be conservative, resulting in missed MEV transactions for several reasons. First, most prior research considers a pair of token transfers as a token exchange only when swap events in DEXs occur. However, token exchanges might occur in a transaction for various reasons beyond DEXs.

In particular, the event \textit{sellGem} in the MakerDAO contract\footnote{0x89B78CfA322F6C5dE0aBcEecab66Aee45393cC5A} is emitted when a MakerDAO vault user locks USDC and takes out a DAI loan. Although this event incurs token exchanges between USDC and DAI, existing algorithms do not treat it as a token exchange. Additionally, not all token exchanges are accompanied by events. For instance, when the functions \textit{leave} and \textit{enter} in the Shiba Inu Staking contract\footnote{0xB4a81261b16b92af0B9F7C4a83f1E885132D81e4} are called, token exchanges between SHIB and xSHIB occur without emitting any events.

Moreover, current methods fail to detect several types of MEV transactions using novel strategies. For example, some MEV transactions involve buying and selling index tokens, such as NFTI (NFT Index) and Crypto Index, which do not form the transfer cycles required for arbitrage transactions as described in \cite{piet2022extracting}. Another example of false-negative errors includes transactions where DeFi services use burn and mint mechanisms when exchanging tokens, as they send tokens to or receive tokens from the null address, not forming a cycle.

While detection algorithm being conservative may exhibit high precision, it performs poorly in recall rate. Consequently, the current method detects only a portion of the total MEV transactions and fails to capture all actual MEV transactions on the Ethereum network.

\section{Proposed method}\label{Proposed Method}
\subsection{Overall method}
Existing methods for MEV detection face three key challenges: dependency, maintenance, and recall rate. Dependency and maintenance problems stem from the need to prepare contract ABIs, relevant function calls and events in advance, while the recall rate issue is primarily due to a lack of understanding of novel MEV strategies.

In our proposed method, we address these challenges through a two-step process: data construction and model construction. In the data construction step, we explore and categorize various MEV types, including MEV transactions that were never discussed before. We classify sandwich attacks into two types and arbitrages into five types, as presented in Table \ref{table:mev_types}. Next, we label transactions as MEV transactions if the belonged to one of the previously categorized types of MEV. This labeled dataset serves as the foundation for training and evaluating our proposed detection model, ArbiNet. 

In the model construction step, we introduce a novel arbitrage detection model, ArbiNet, for detecting arbitrages and an heuristic algorithm for detecting sandwich attacks. Our approaches simplify the data construction process by removing the need for DeFi contract ABI or event information. Only ERC-20 token transfer data is required, which can be easily extracted from transactions due to the unchanging nature and easy access of the ERC-20 ABI.


\begin{table}[h!]
\caption{Types of MEV Transaction Forms}
\begin{tabular}{ccl}
\toprule
Category & Type & Description \\
\midrule
\multirow{2}{*}{Sandwich} & $S1$ & Single DEX Sandwich  \\
& $S2$ & Cross-DEX Sandwich \\
\hline
\multirow{5}{*}{Arbitrage} & $A1$ & Simple Loop Arbitrage \\
& $A2$ & Burn \& Mint Mechanism Arbitrage \\
& $A3$ & Set Token Arbitrage \\
& $A4$ & Multi Address Arbitrage \\
& $A5$ & NFT Arbitrage \\
\hline
\end{tabular}
\label{table:mev_types}
\end{table}

\begin{figure*}[h]
  \centering
  \includegraphics[width=\linewidth]{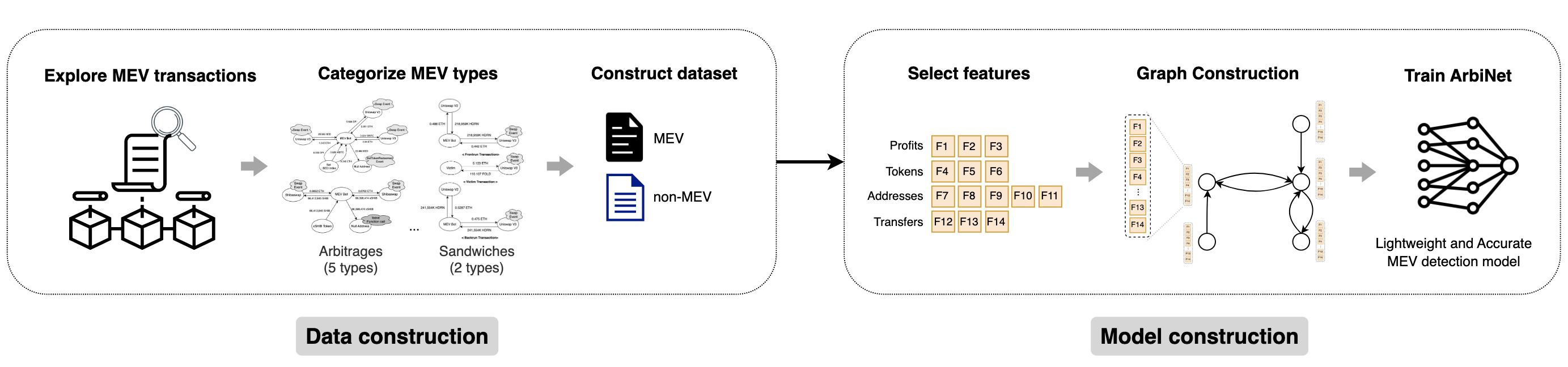}
  \caption{Overall Method}
  \Description{Overall Method}
  \label{fig:Overall Method}
\end{figure*}

From the perspective of a user using our method, only transaction receipts and ERC-20 ABI are necessary, as they can utilize the sandwich detection algorithm and the pretrained ArbiNet model. We make our models publicly available to facilitate their ease of use and adoption. Figure \ref{fig:Overall Method} provides a summary of our overall method.

\subsection{Model}
\subsubsection{Sandwich Detection Algorithm} \hfill

\noindent 
We present a simple algorithm for detecting sandwich transactions, as outlined in Algorithm \ref{alg:sandwich detection algorithm}. Unlike previous methods using contract data, the algorithm operates solely on token transfer data, aiming to identify pairs of sandwich transactions that have generated profits. 

The algorithm initiates by examining pairs of transactions sharing the same recipient address within a single block. Subsequently, it filters out transaction pairs that do not satisfy specific criteria, which are indicative of non-sandwich transactions. For example, the filtering conditions encompass cases where the traded token count is less than 1, or the cumulative profits derived from both transactions result in a negative value.

\SetKwComment{Comment}{/* }{ */}
\begin{algorithm}[h]
\SetAlgoLined
\KwIn{${tx}_1,{tx}_2,...,{tx}_n$}
\KwOut{${sandwich}\_list$}
Initialization\;
$sandwich\_list = []$;

\For{$i \leftarrow 1$ \KwTo $n$}{
  1. Find ${tx}_j$ which satisfies ${tx}_i.to = {tx}_j$.to;
  
  2. ${profits}_i = calculate\_profits({tx}_i)$;\\
  \tcc*[r]{Dictionary token(key):amount(value)}
  3. ${profits}_j = calculate\_profits({tx}_j)$;

  4. $Total\_profits = {profits}_i + {profits}_j$;

  5. \textbf{if} $len({profits}_i) \leq 1 $ or $len({profits}_j) \leq 1:$ \hspace{11cm} \textbf{continue}
  
  6. \textbf{if} $val \geq 0$ \textbf{for} $token,val$ \textbf{in} $ {profits}_i.items()$ \hspace{11cm} \textbf{continue}

  7. \textbf{if} $val == 0$ \textbf{for} $token, val$ \textbf{in} $ {profits}_j.items()$ \hspace{11cm} \textbf{continue}

  8. \textbf{if} ${profits}_i.keys() \not\subset {profits}_j.keys()$ \textbf{or} ${profits}_j.keys() \not\subset {profits}_i.keys()$  \hspace{11cm} \textbf{continue}

  9. \textbf{if} $val < 0$ \textbf{for} $token, val$ \textbf{in} ${Total\_profits}.items()$    \hspace{11cm} \textbf{continue}

  $sandwich\_list += [{tx}_i, {tx}_j]$
}
\caption{Sandwich Detection}
\label{alg:sandwich detection algorithm}
\end{algorithm}

\subsubsection{ArbiNet : Arbitrages Detection Model} \hfill

\begin{figure*}[h]
  \centering
  \includegraphics[width=\linewidth]{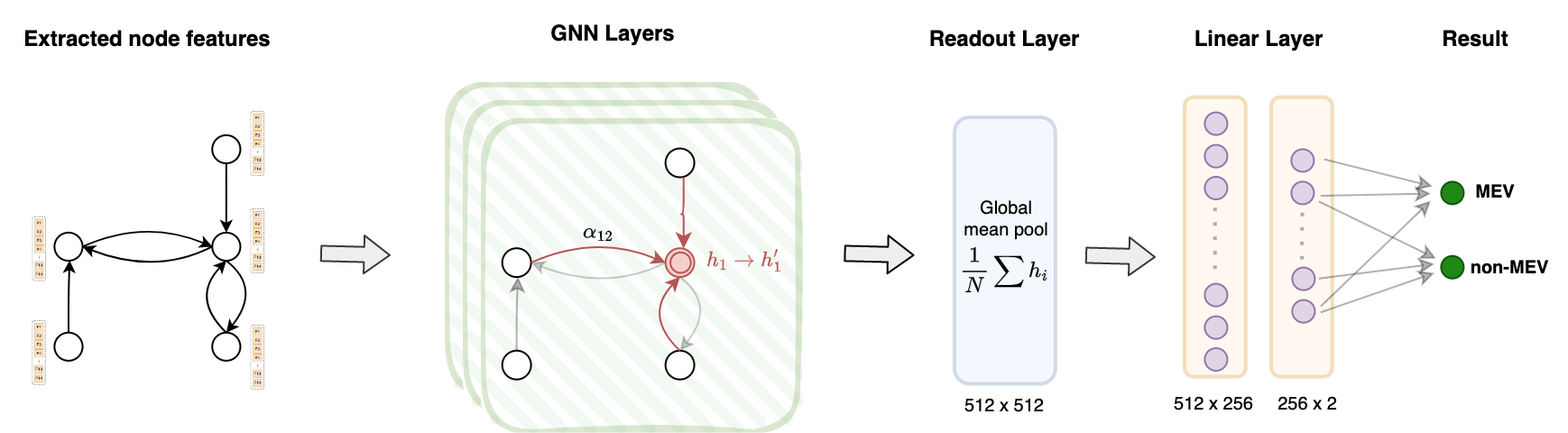}
  \caption{ArbiNet}
  \Description{ArbiNet}
  \label{fig:ArbiNet}
\end{figure*}

\noindent
Arbitrage transactions often exhibit more complex forms than sandwich transactions, making it challenging for simple heuristics to accurately identify and capture them. To address this, we propose ArbiNet, a Graph Neural Network (GNN)-based arbitrage detection model that relies solely on ERC-20 token transfer data as input. The overall structure of ArbiNet, shown in Figure \ref{fig:ArbiNet}, comprises GNN layers including Graph Convolutional Networks (GCN), GraphSAGE, and Graph Attention Networks (GAT).

For each transaction, we first extract token transfer data and construct a graph by extracting features from each node in the associated token transfer graph. We initially extract 14 features for each node in a token transfer graph. This constructed graph serves as input for the GNN layers, which compute hidden states for the node features.

Following the GNN layers, the readout layer calculates the global mean of the hidden states across all nodes. A linear layer then processes the output from the readout layer and computes the binary classification, sending the results to two output nodes representing the MEV and non-MEV classes.

To prepare the training data for ArbiNet, we labeled arbitrage transactions as 1 and non-arbitrage transactions as 0. We collected labeled arbitrages based on the five types of arbitrage transaction forms we explored.

\noindent \textbf{Why GNN?} 
We chose Graph Neural Networks (GNN) for our model because they are efficient in handling graph data. One crucial insight that can be derived from analyzing ERC-20 token transfer data is the presence of loops within token transfers, which can be indicative of potential arbitrage opportunities. Edge directions are important, but many machine learning models, such as Support Vector Machines \cite{pisner2020support} and logistic Regression \cite{hosmer2000applied}, cannot handle relationships between edges effectively.

A potential alternative approach would be to implement a code-based solution for detecting loops and calculating profit within a transaction, thus identifying arbitrage transactions. However, this method is not feasible for two main reasons:

First, diverse forms of arbitrage transactions would require individual hand-written code for each form. For example, there are arbitrage transactions that do not form token loops and use different addresses for burning and minting tokens, depending on the contracts they interact with.

Second, the logic would need regular updates to accommodate new types of arbitrage transactions, necessitating continuous observation and adaptation.

Whereas the simple heuristic needs constant upkeep and requires individual handling of arbitrage strategies, GNNs can learn typical patterns observed in arbitrage transactions. They efficiently aggregate information from neighboring nodes to create node representations. By utilizing these representations and labels, the model can learn unique patterns shared by arbitrage transactions. Overall, GNNs offer better maintainability since they do not require humans to specify the exact forms of each type.

\noindent \textbf{Graph Construction}
Previous research on MEV detection using neural networks \cite{varun2022mitigating} has primarily focused on meta transaction data, such as gas prices. While informative, this approach may not fully capture the graph structure of token transfers during transaction execution, potentially limiting the model's performance.

Existing GNN research in Ethereum \cite{liu2022fa, tan2021graph} has mainly targeted anomaly detection, often using graphs with a single type of edge. Our work is the first to leverage graphs with multiple token types, utilizing the rich information available in token transfer data. Additionally, our research is the first to apply GNNs to arbitrage detection, broadening the scope and applicability of GNN models in Ethereum transaction analysis.

The use of GNN layers on ERC-20 token transfer graphs poses challenges due to multiple edge types, each representing a different type of token transfer. Heterogeneous Graph Neural Networks (HGNN), such as Relational Graph Convolutional Networks (R-GCNs)\cite{schlichtkrull2018modeling} and Heterogeneous Graph Attention Networks (HANs)\cite{wang2019heterogeneous}, are designed to handle multiple edge types. However, these models assume a fixed and known number of categorical edges or edge types. They are not applicable to ERC-20 token graphs, as the edge types and the total number of types are unknown.

To address this issue, we develop a method for constructing graphs that can be easily used as input for widely known GNN layers, such as GCN and GAT layers. Our constructed graph can be obtained from token transfer graphs effortlessly and extracts information relevant to the classification of arbitrage transactions.

\noindent \textbf{Feature Selection}
For input node features of ArbiNet, we selected 14 features for each node, which are considered crucial in identifying arbitrage transactions. These features can be divided into four groups based on the information they represent: profits, tokens, addresses, and transfers. Table \ref{table:Node Features} summarizes the selected features.

(1) \textit{Feature 1 \textasciitilde  Feature 3 (Profits)}: 
Feature 1 represents the count of tokens with profits less than zero. Feature 2 and Feature 3 denote the count of tokens with a profit greater than 0 and equal to 0, respectively. These features are essential in detecting arbitrage taker's addresses, as most arbitrage transactions involve multiple tokens with positive profits and no tokens with negative profits.

(2) \textit{Feature 4 \textasciitilde Feature 6 (Tokens) }:
The number of distinct tokens sent and received by each address (Features 4 and 5) is valuable for detecting arbitrages, as DEX addresses typically send one token and receive another during token exchanges. By aggregating token counts from neighboring nodes, the model gains useful insights about different tokens involved in a transaction. Feature 6 provides the total number of tokens traded in an entire transaction, further enhancing the model's ability to derive valuable insights.

(3) \textit{Feature 7 \textasciitilde Feature 11 (Addresses) }:
The null address (0x00...00) is commonly used for burning tokens or minting new tokens (Feature 7). It is often observed in two types of arbitrage transactions, $A2$ and $A3$. This feature allows the model to learn about the burn and mint mechanisms involved in these transactions.

Feature 8, indicating whether an address is the builder of the block, provide insights into potential arbitrage opportunities. Many MEV transactions offer private tips to the block builder to increase the likelihood of inclusion in the blocks. Feature 8 also enables the model to differentiate between tips and profits resulting from arbitrages.

In most MEV transactions, the MEV taker is a Contract Address (CA) deployed to call multiple DeFi contracts within a single transaction (Feature 9). Additionally, the MEV taker often transfers the realized profits to their own Externally Owned Accounts (EOA). Knowing whether an address is an EOA or CA aids in classifying MEV transactions.

Feature 10 indicates if an address is the sender of the transaction, which is important because if a transaction is an MEV transaction, the sender address also belongs to the MEV taker, and the gross profits of the sender address should be considered.

Feature 11 identifies if an address is the recipient of the transaction, providing clues about whether an address is a contract address associated with the MEV taker.

(4) \textit{Feature 12 \textasciitilde Feature 14 (Transfers) }:
Feature 12, 13 and 14 represent information related to the count of transfers. While these features may exhibit some correlation with Features 4, 5, and 6, the number of transfers often contains valuable insights that can help infer the type of nodes and edges involved. For instance, a node that demonstrates a high count of transfers but a low count of distinct token types is likely to be a aggregation router service.

\begin{table}[h!]
\caption{Node Features}
\begin{tabular}{ccp{5.3cm}}
\toprule
Group & Feature & Description \\
\midrule
\multirow{3}{*}{Profits} & F1&\# of tokens whose profit is smaller than 0 \\
&F2&\# of tokens whose profit is greater than  0\\
&F3&\# of tokens whose profit is 0\\
\hline
\multirow{3}{*}{Tokens} &F4&\# of tokens sent at least once at the address\\
&F5&\# of tokens received at least once at the address\\
&F6&\# of tokens transferred at least once in this transaction\\
\hline
\multirow{5}{*}{Addresses} &F7&If the address is null address or not\\
&F8&If the address is builder address or not\\
&F9&If the address is CA or EOA \\
&F10&If the address is from address of the transaction or not \\
&F11&If the address is to address of the transaction or not\\
\hline
\multirow{3}{*}{Transfers} &F12&\# of transfers sent from the address\\
&F13&\# of transfers received at the address\\
&F14&\# of transfers in this transaction\\
\hline
\end{tabular}
\label{table:Node Features}
\end{table}

\section{In-depth exploration of MEV transactions}\label{classification}
In this section, we focus on the data construction step of our proposed method, as described in Section \ref{Proposed Method}. We first provide an in-depth exploration of MEV transactions and classify them based on their forms and strategies. With the classifications established, we label MEV transactions using smart contract events and function calls that we obtained manually.


A crucial aspect of MEV detection is identifying intended token exchanges. Current algorithms \cite{piet2022extracting,qin2022quantifying,flashbots-mevispectpy} predominantly concentrate on a limited set of events and function calls as intended token exchange actions, primarily tracking swap events in DEXs. However, numerous DeFi protocols, besides DEXs, now support various forms of token exchanges. For instance, LSD (Liquid Staking Derivatives) services like Lido send stETH to users who deposit their ETH. We have identified 47 events and 18 function calls that trigger intended token exchanges, which are organized in Appendix Table \ref{tab:token exchange events} and \ref{tab:token exchange functions}.

In addition to the diverse forms of token exchanges and events that existing algorithms often overlook, our study aims to provide an in-depth exploration of MEV strategies, addressing the limited forms captured by heuristics employed by previous works. Despite the conceptual definitions for each type of MEV transaction introduced by \cite{weintraub2022flash}, they are not directly applicable to MEV detection due to the variability and numerous exceptions in the forms of MEV transactions.

To the best of our knowledge, we are the first to categorize MEV transactions based on their forms and strategies, including novel ones. We present 2 types of strategies in sandwich transactions and 5 types of arbitrage transactions.

\subsection{Sandwich}

\subsubsection{S1 : Single DEX Sandwich}
\begin{figure}[h]
  \centering
  \includegraphics[width=0.8\linewidth]{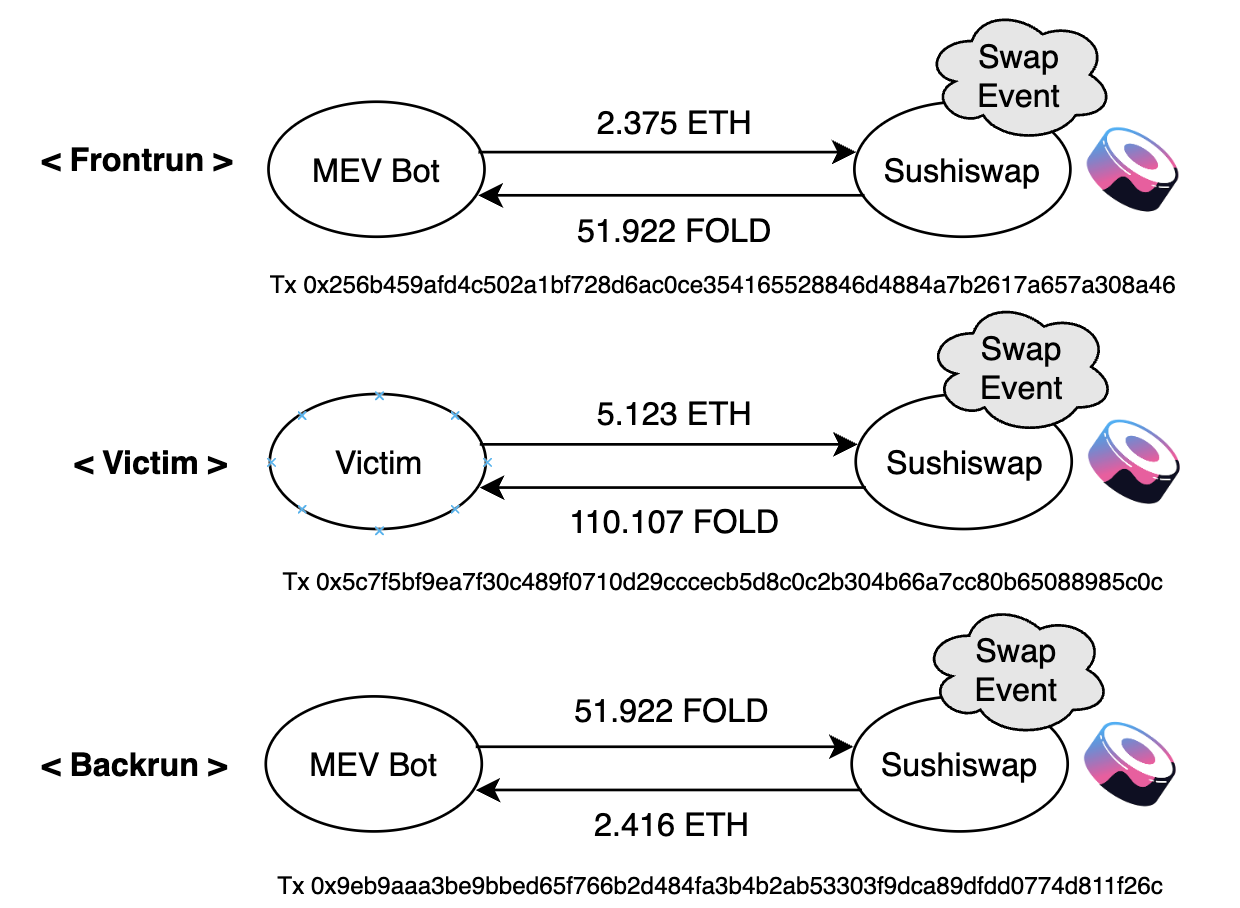}
  \caption{Sandwich Transaction Example 1}
  \Description{Sandwich Transaction Example 1}
  \label{fig:Sandwich Example1}
\end{figure}

The Single DEX Sandwich (S1) is the most basic form of a sandwich transaction, and it is the primary focus of existing algorithms. In this type of transaction, when an MEV taker identifies a victim transaction that swaps token $A$ for token $B$ on a single decentralized exchange (DEX), referred to as ${DEX}_1$, the taker front-runs the victim transaction by swapping token $A$ for token $B$ first. As a result, the victim's transaction swaps A for a smaller amount of B than initially expected. Finally, the MEV taker uses token $B$ received in the front-running transaction to swap $B$ for $A$, realizing a profit. This profit comes from the victim's unexpected loss, as shown in Figure \ref{fig:Sandwich Example1}.

\subsubsection{S2 : Cross-DEX Sandwich}
\begin{figure}[h]
  \centering
  \includegraphics[width=0.8\linewidth]{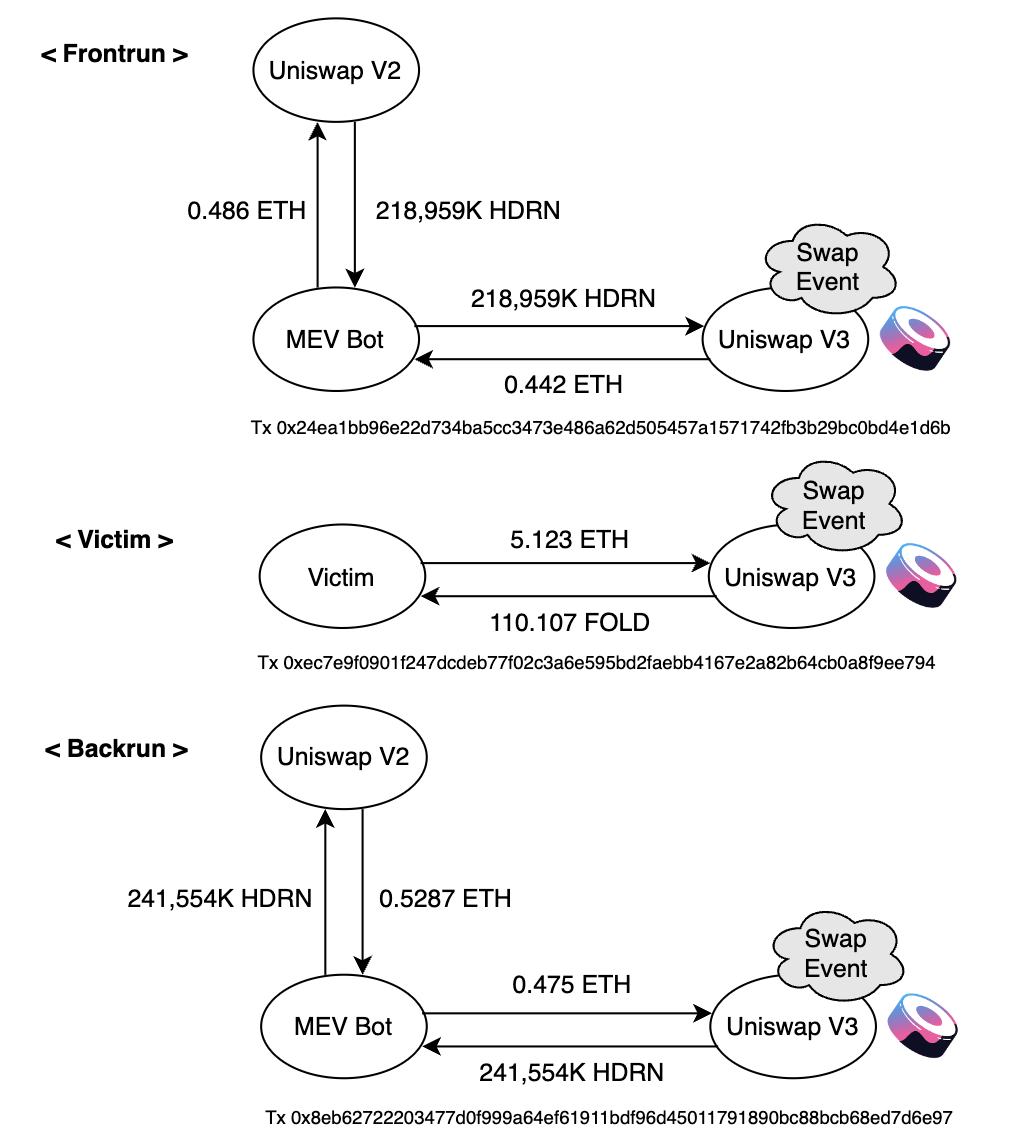}
  \caption{Sandwich Transaction Example 2}
  \Description{Sandwich Transaction Example 2}
  \label{fig:Sandwich Example2}
\end{figure}

The Cross-DEX Sandwich ($S2$) covers cases where the MEV taker does not have token A but possesses token B in advance. Unlike the Single DEX Sandwich ($S1$), the taker initially swaps token B for token A on another DEX, called ${DEX}_2$, before the victim's transaction. After the victim's transaction, the taker swaps B for A on ${DEX}_1$ and then swaps $A$ for $B$ on ${DEX}_2$. Figure \ref{fig:Sandwich Example2} illustrates an example of this type of transaction.

In the front-running transaction, the taker experiences a net loss, as they buy token B on ${DEX}_2$ and sell B at a lower price on ${DEX}_1$. However, in the back-running transaction, the taker earns profits from two sources. One source is the sandwich profit resulting from the victim's loss, and the other is the arbitrage profit between ${DEX}_1$ and ${DEX}_2$. The taker recovers the value lost in the front-running transaction by buying back token B at a lower price. Interestingly, the front-running transaction resembles a failed arbitrage attempt, while the back-running transaction appears as a highly successful arbitrage transaction.

We note that the existing sandwich detection method (\cite{weintraub2022flash}, \cite{qin2022quantifying}) cannot capture Type 2 sandwich transaction, because the amount of token sold in the frontrun transaction differs from the amount in the backrun transaction. 

\subsection{Arbitrage}

\subsubsection{A1 : Simple Loop Arbitrage}
\begin{figure}[h]
  \centering
  \includegraphics[width=0.7\linewidth]{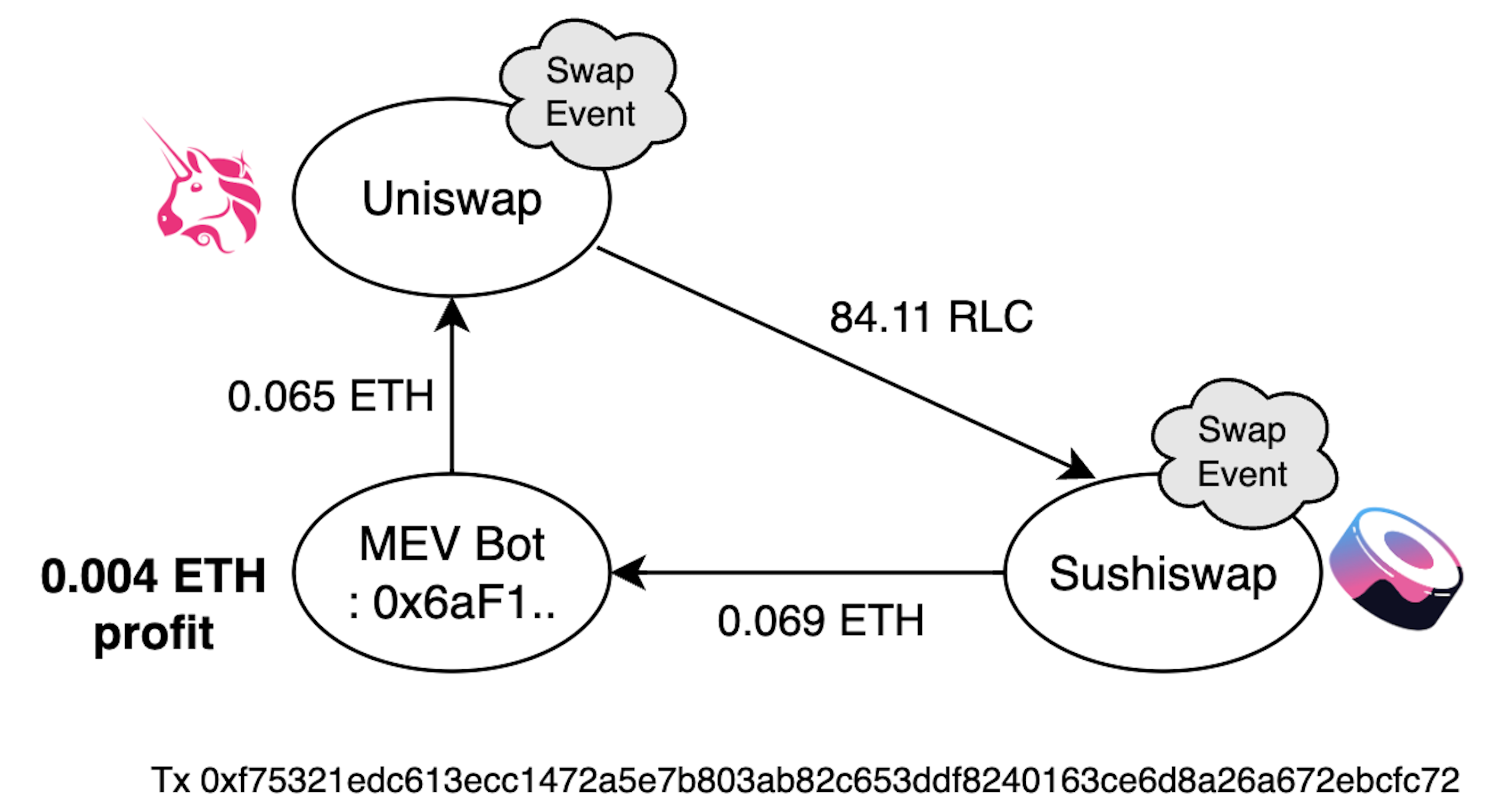}
  \caption{Arbitrage Type 1 transaction}    
  \Description{Arbitrage Type 1 transaction}
  \label{fig:Arbitrage1}
\end{figure}

Type 1 arbitrage transactions represent the most common form of arbitrage. The MEV taker exchanges a token within a DeFi protocol and subsequently trades the received token in another protocol at a higher price. Figure \ref{fig:Arbitrage1} illustrates an example of a Type 1 arbitrage transaction. Loops are formed by defined token exchanges, resulting in a net positive profit for a single MEV taker address. In accordance with the style employed by \cite{piet2022extracting} and \cite{qin2022quantifying}, Type 1 arbitrage transactions do not include pairs of adjacent edges without any associated events.

\subsubsection{A2 : Burn \& Mint Mechanism Arbitrage}
In numerous transactions, tokens are exchanged using burn and mint mechanisms. DeFi protocols,such as Liquid Staking Derivatives (LSD) protocols, employ this method by burning tokens sent to a null address and minting new tokens issued from a different address.

\begin{figure}[h]
  \centering
  \includegraphics[width=\linewidth]{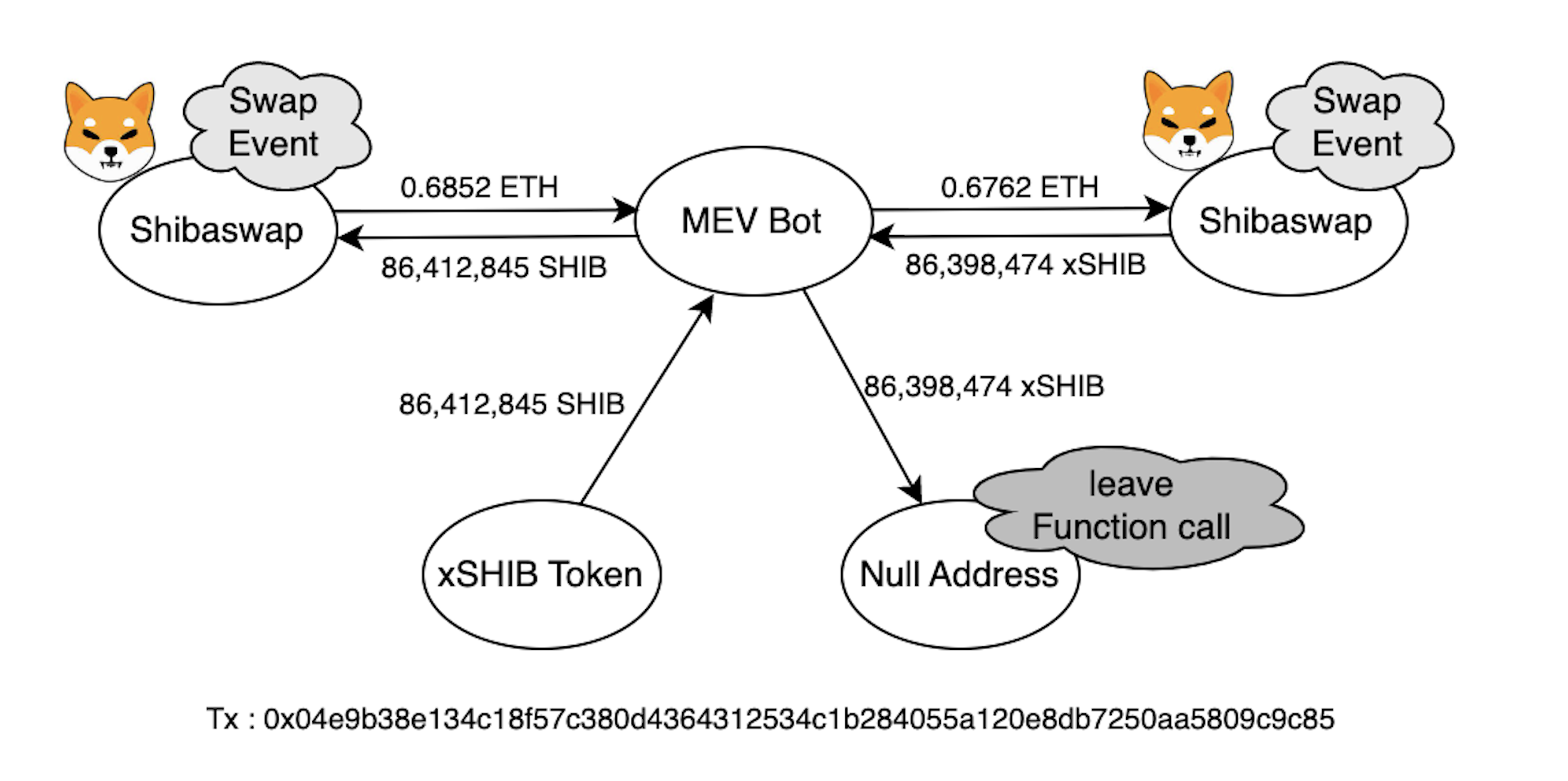}
  \caption{Arbitrage Type 2 transaction}
  \Description{Arbitrage Type 2 transaction}
  \label{fig:Arbitrage2}
\end{figure}

As illustrated in Figure \ref{fig:Arbitrage2}, the MEV taker executes two swaps: one for converting ETH to xSHIB and the other for trading SHIB back to ETH. Additionally, the taker exchanges xSHIB for SHIB by invoking the "leave" function in the xSHIB contract, without emitting any events. We observe that existing algorithms fail to detect Type 2 transactions for two primary reasons. First, token transfer graphs do not create a cycle, as the MEV taker sends xSHIB to a null address but receives SHIB from the xSHIB contract address. Second, since there is no event emitted, tracking over events miss this type of token exchange.

Type 2 arbitrage underscores the difficulty of achieving perfect arbitrage detection, given that event logs do not capture all token exchanges and that sender and receiver addresses do not always correspond to the token contract address.

\subsubsection{A3 : Set Token Arbitrage}
\begin{figure}[h]
  \centering
  \includegraphics[width=\linewidth]{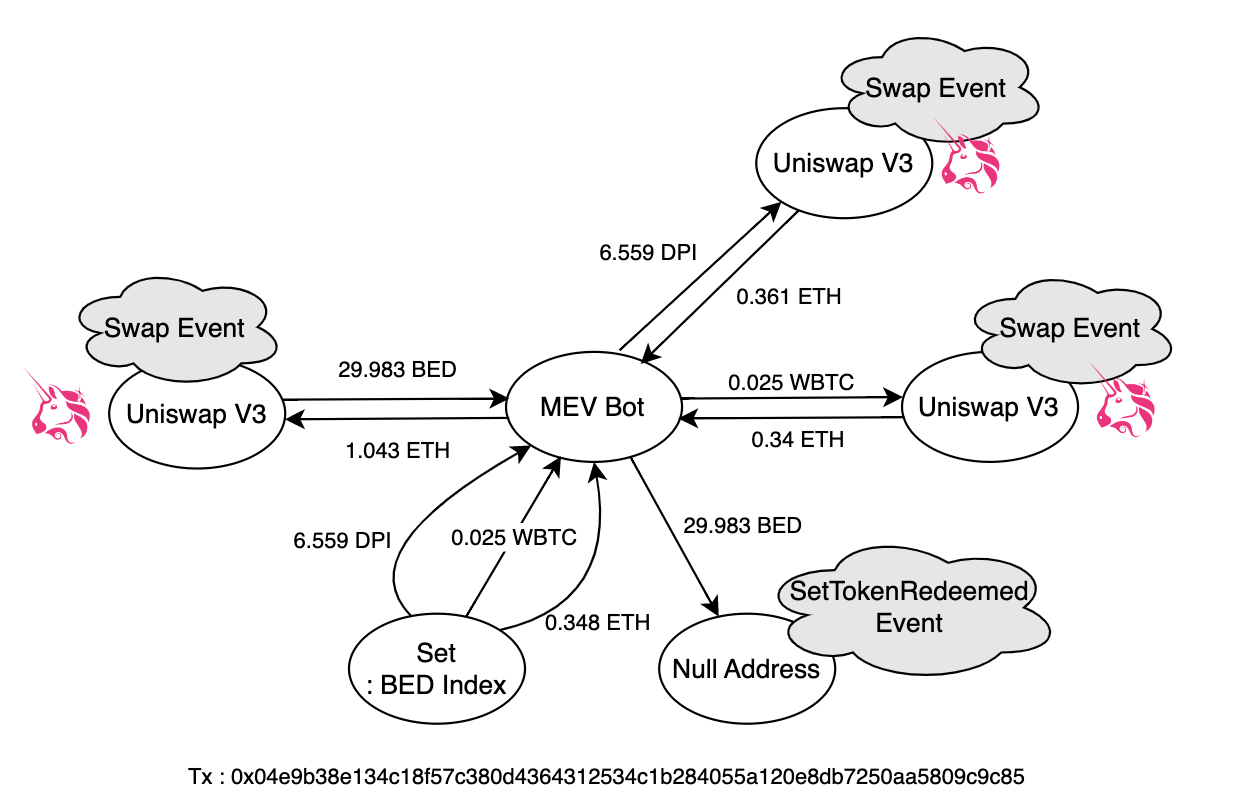}
  \caption{Arbitrage Type 3 transaction}
  \Description{Arbitrage Type 3 transaction}
  \label{fig:Arbitrage3}
\end{figure}

Type 3 arbitrage transactions involve profiting from set tokens. A set token (\textit{e.g.} Crypto Index, NFT Index) represents a portfolio of underlying assets. In Figure \ref{fig:Arbitrage3}, the BED Index token comprises BTC, ETH, and DPI (DeFiPulse Index). The MEV taker initially purchases BED on Uniswap, then redeems DPI, BTC, and ETH by burning their BED. Subsequently, the taker swaps DPI and BTC for ETH, ultimately generating a profit of 0.005 ETH.

Traditional detection methods are also ineffective in identifying Type 3 arbitrage transactions, as they do not form any discernible loops. This is due to the characteristic of set tokens that, in contrast to swap events, redeeming set tokens usually yields more than one asset.

\subsubsection{A4 : Multi Address Arbitrage}
\begin{figure}[h]
  \centering
  \includegraphics[width=0.8\linewidth]{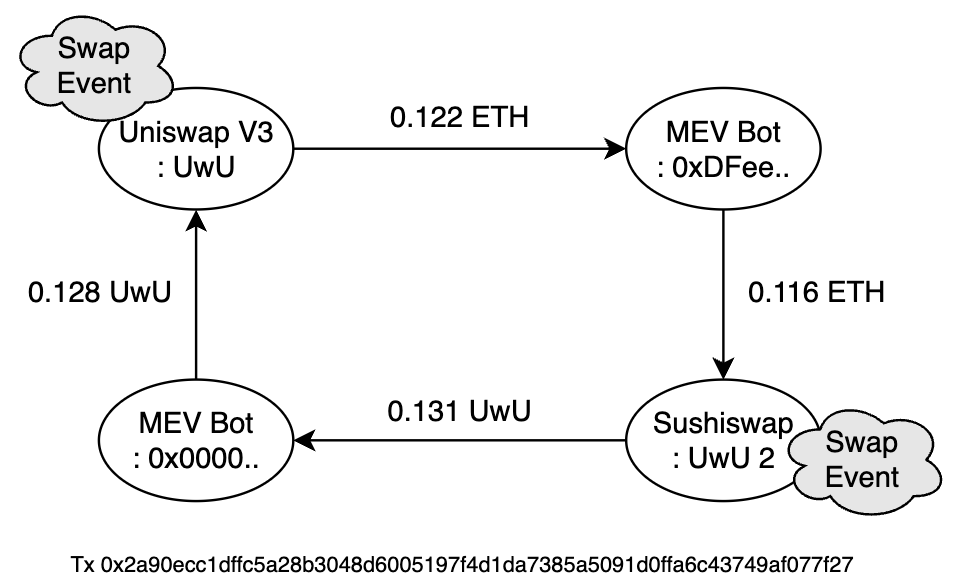}
  \caption{Arbitrage Type 4 transaction}
  \Description{Arbitrage Type 4 transaction}
  \label{fig:Arbitrage4}
\end{figure}

In Type 4 arbitrage transactions, an MEV taker utilizes two addresses within a single arbitrage transaction. As demonstrated in Figure \ref{fig:Arbitrage4}, Type 4 arbitrage involves two addresses, resulting in profit generated in both addresses. In this case, the total profit should be the sum of profits from each address. We consider an address as the MEV taker's second address when the amount of tokens sent from the address is smaller than the amount of tokens received and the tokens are the same.

\subsubsection{A5 : NFT Arbitrage}

As NFT exchanges (\textit{e.g.} Opensea, Blur) gain traction, NFT trading volume has surged. Platforms like Sudoswap enable instant swaps between NFTs and ERC-20 tokens, leading to emerging arbitrage opportunities involving NFTs. In this paper, we highlight a novel NFT-based arbitrage strategy, contributing to the understanding of innovative arbitrage possibilities in the space.

\begin{figure}[h]
  \centering
  \includegraphics[width=0.8\linewidth]{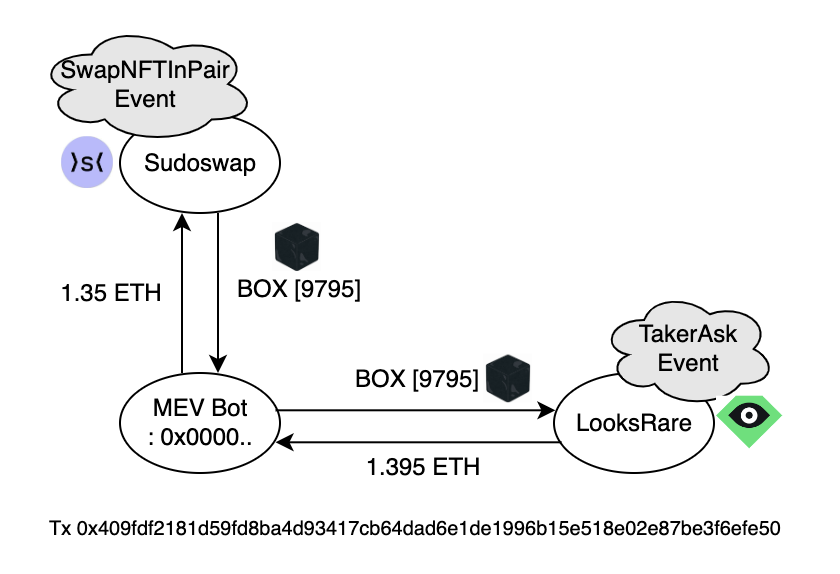}
  \caption{Arbitrage Type 5 transaction}
  \Description{Arbitrage Type 5 transaction}
  \label{fig:Arbitrage5}
\end{figure}

Type 5 arbitrage transactions capture profits generated with NFTs. Figure \ref{fig:Arbitrage5} presents an example of a Type 5 arbitrage transaction. In Sudoswap, the MEV taker swaps a Black BOX NFT with token ID 9795 for 1.35 ETH and then sells the NFT on the LooksRare exchange for 1.395 ETH, resulting in a profit of 0.045 ETH.

Type 5 arbitrage transactions resemble Type 1 transactions, with the exception that ERC-721 tokens are bought and sold instead. To maintain the simplicity of our model, we excluded ERC-721 transfer data when training the arbitrage detection model. Instead, we developed a simple algorithm for detecting Type 5 arbitrage transactions in Algorithm \ref{alg:NFT Arbitrage detection algorithm}.

\section{Evaluation and Analysis}\label{evaluation}

\subsection{Labeled MEV Comparison}
To the best of our knowledge, there are two public datasets that present actual MEV transactions. \cite{weintraub2022flash} provides actual arbitrage data from block 10,000,000 to 14,444,725, and ZeroMEV\cite{ZeroMEV} displays real-time MEV data using Flashbots API\cite{flashbots-mevispectpy}. To validate the completeness of the data used for training and evaluation of ArbiNet, we compared the total number of labeled transactions in our dataset to the numbers reported in the two public datasets.

For the same Ethereum block range as \cite{weintraub2022flash} (10,000,000 to 14,444,725), our labeled data contained 409,467 arbitrage transactions, 20.6\% higher than the 339,477 transactions in the public dataset. When compared to Flashbots data from block 15,500,000 to 15,650,000, our dataset identified 121,543 arbitrage and 67,148 sandwich transactions, while Flashbots reported 63,492 arbitrage and 63,877 sandwich transactions, resulting in a 91.4\% difference in arbitrage and a 5.1\% difference in sandwich transactions.

\begin{figure}[h]
  \centering
  \includegraphics[width=0.8\linewidth]{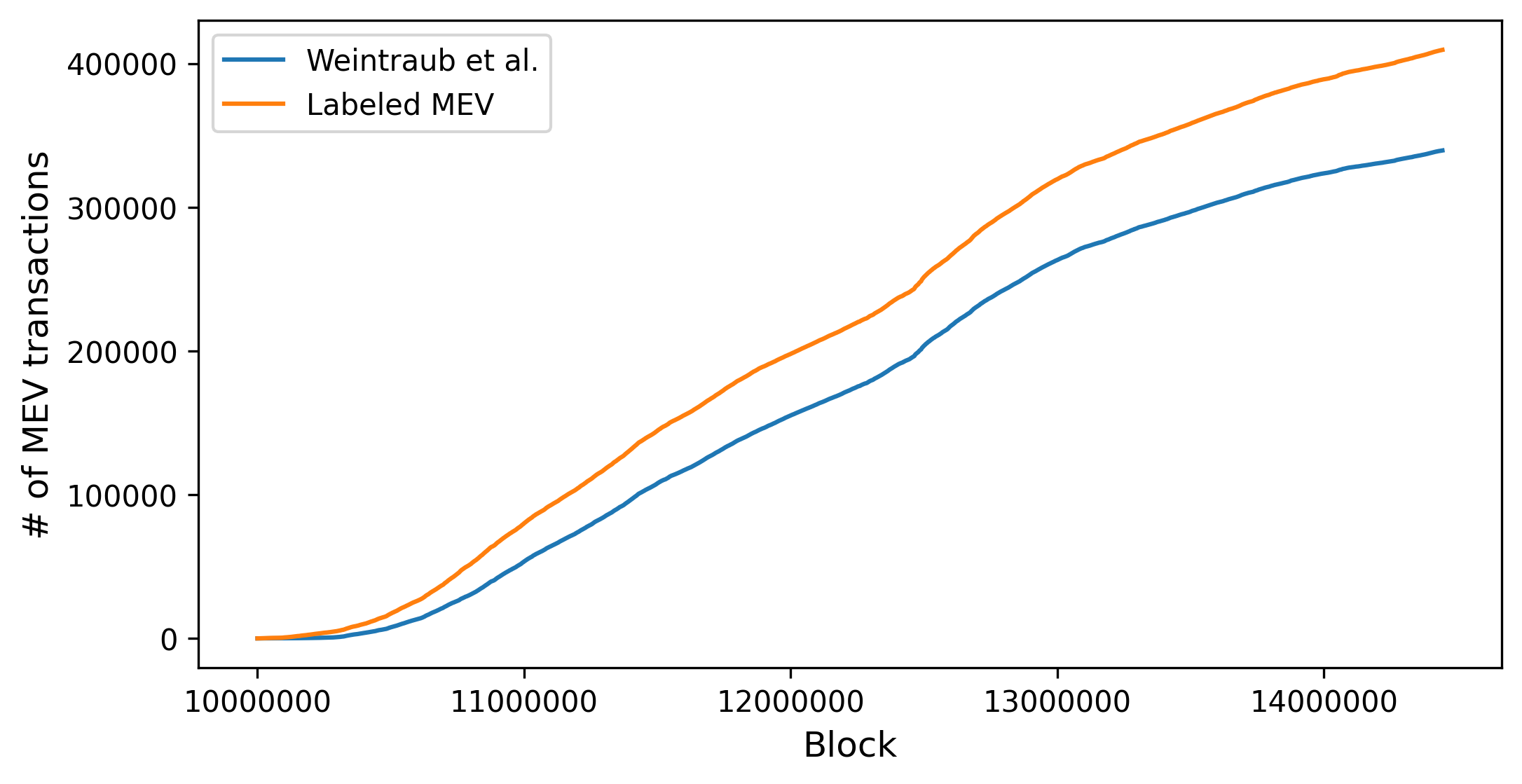}
  \caption{Labeled MEV vs Weintraub et al.}
  \label{fig:pan_comparison}
\end{figure}

\begin{figure}[h]
  \centering
  \includegraphics[width=0.8\linewidth]{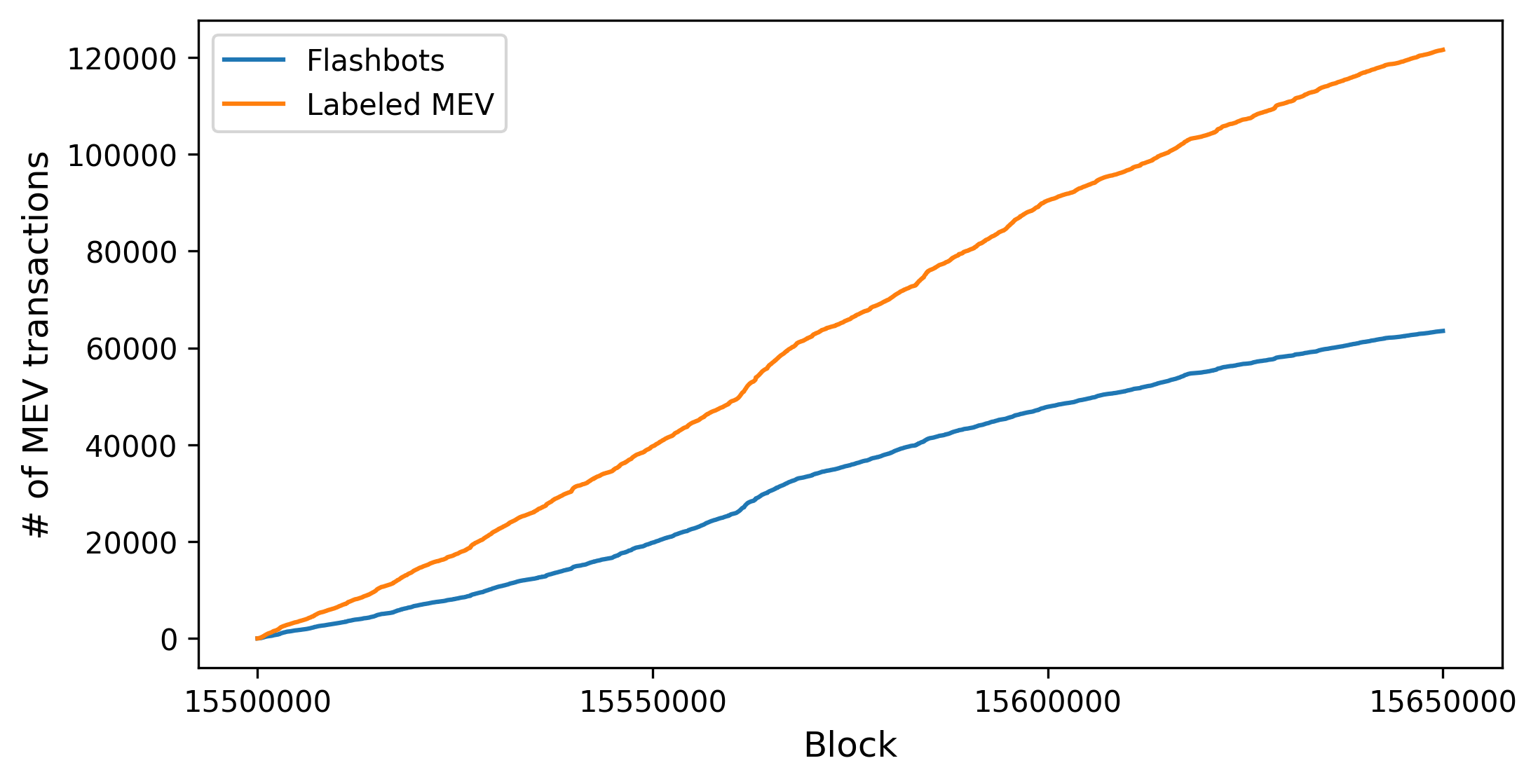}
  \caption{Labeled MEV vs Flashbots}
  \label{fig:flashbots_comparison}
\end{figure}

These comparisons, supported by Figure \ref{fig:pan_comparison} and Figure \ref{fig:flashbots_comparison}, demonstrate that our labeled MEV transaction dataset surpasses public data sources, offering a more comprehensive representation of the full range of MEV transactions.

\subsection{Smart Contract Diversification in MEV}
\begin{figure}[h]
  \centering
  \includegraphics[width=0.8\linewidth]{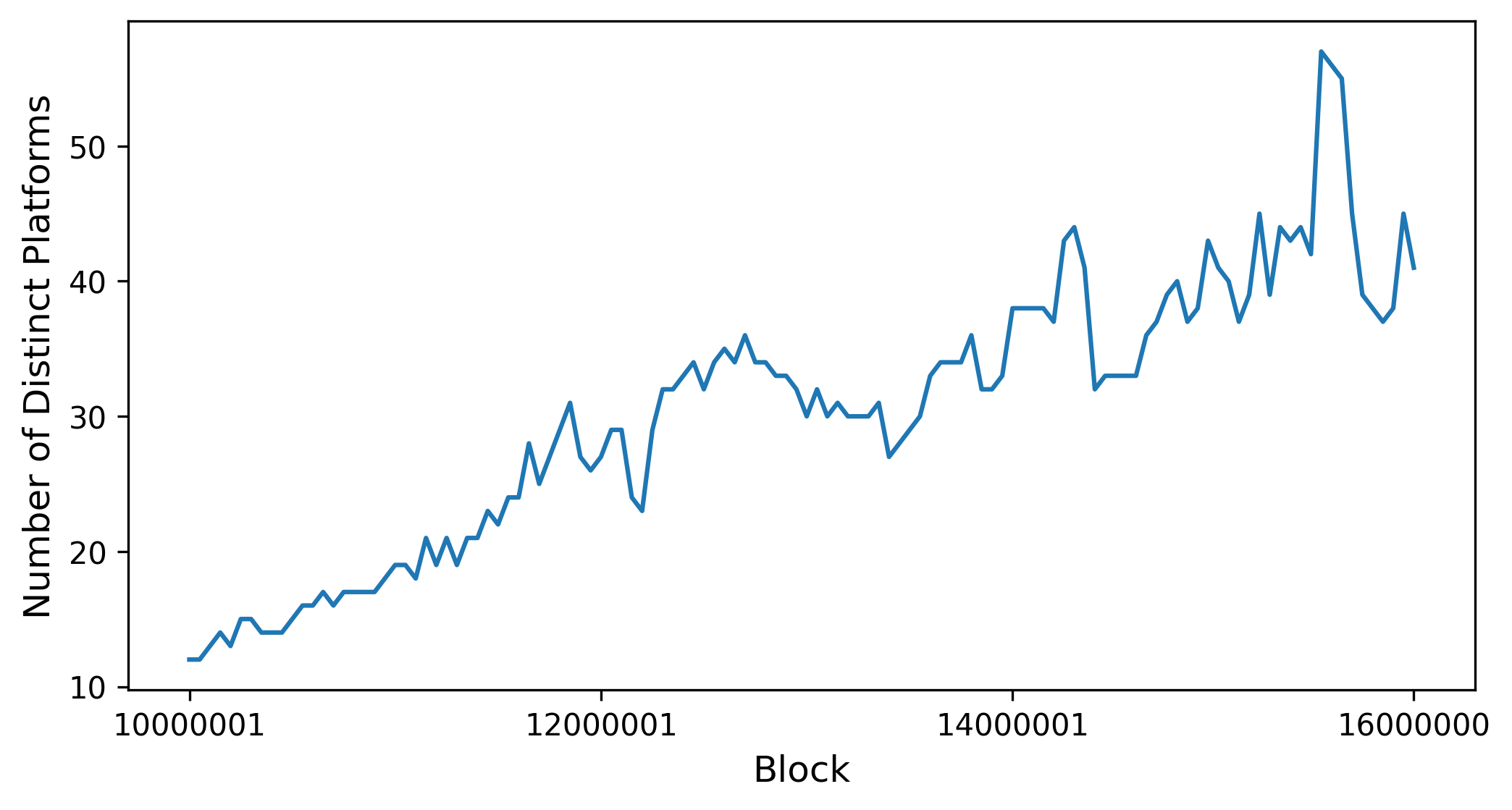}
  \caption{Count of Smart Contracts in MEV transactions}
  \label{fig:Contracts Count}
\end{figure}

The forms and strategies of MEV transactions are not only growing, but the smart contracts utilized in MEV transactions are also increasing and diversifying. To assess these changes, we investigated the number of distinct smart contracts involved in MEV transactions. We analyzed data from block 10,000,000 to 16,000,000 and plotted the counts of smart contracts used in MEV transactions. The results in Figure \ref{fig:Contracts Count}, indicate a general trend of a growing number of contracts being employed to profit from market inefficiencies.

\begin{figure}[h]
  \centering
  \includegraphics[width=\linewidth]{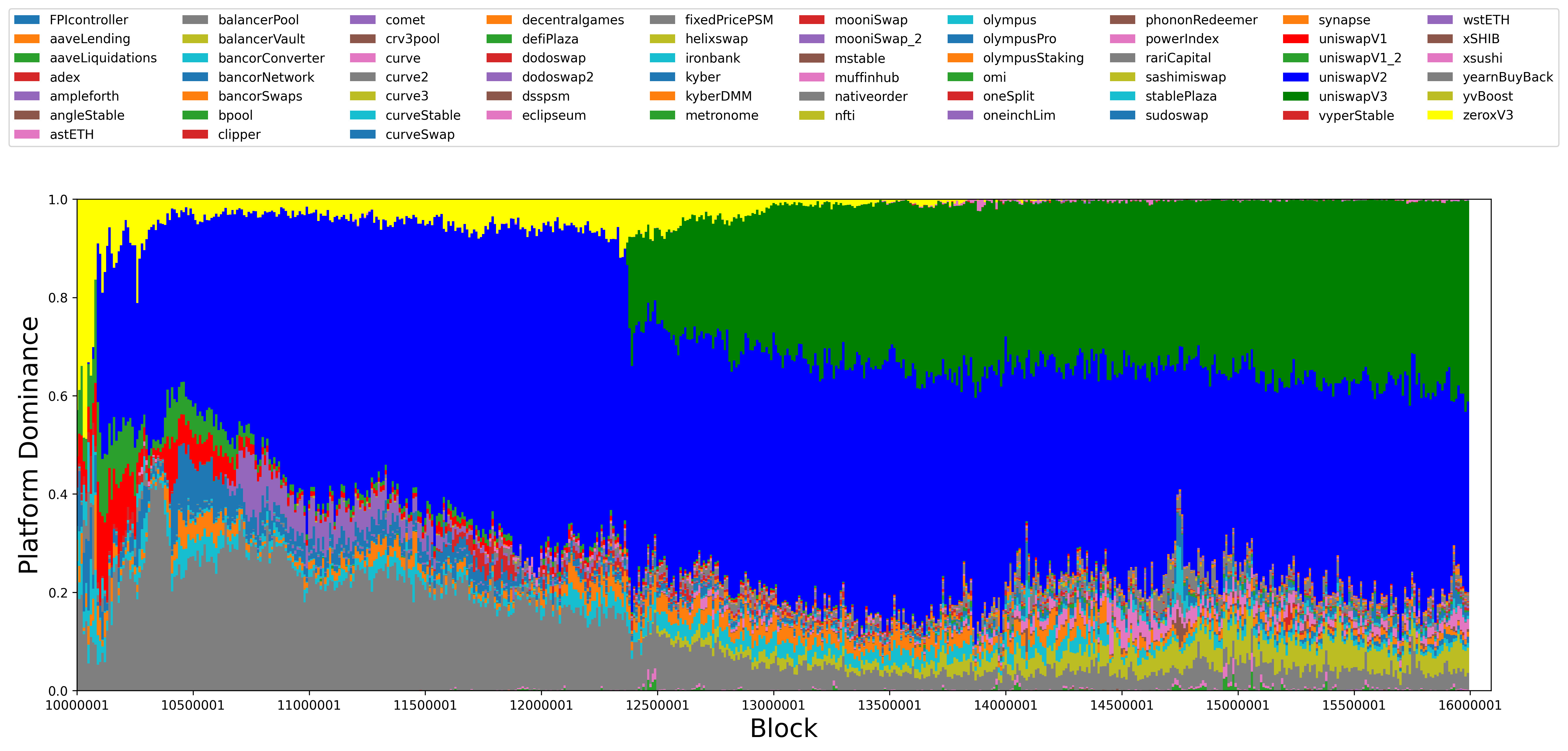}
  \caption{Dominance of each smart contract in MEV transactions}
  \label{fig:Contracts Ratio2}
\end{figure}

Using the same data, we plotted the dominance of each smart contract used in MEV transactions every 20,000 blocks. For instance, if two MEV transactions utilized contracts $C_X$ and $C_Y$, $C_X$ and $C_X$ respectively, the ratio of $C_X$, $C_Y$, $C_Z$ would be 0.5, 0.25, and 0.25. The results in Figure \ref{fig:Contracts Ratio2} demonstrate that Uniswap V2, depicted in blue, accounts for nearly half of all MEV transactions. Since its launch around block 12,500,000, the ratio of Uniswap V3 has consistently grown. Conversely, the ratios of 0x V3 and Balancer Pool, represented in yellow and gray, have substantially decreased.

\begin{figure}[h]
  \centering
  \includegraphics[width=\linewidth]{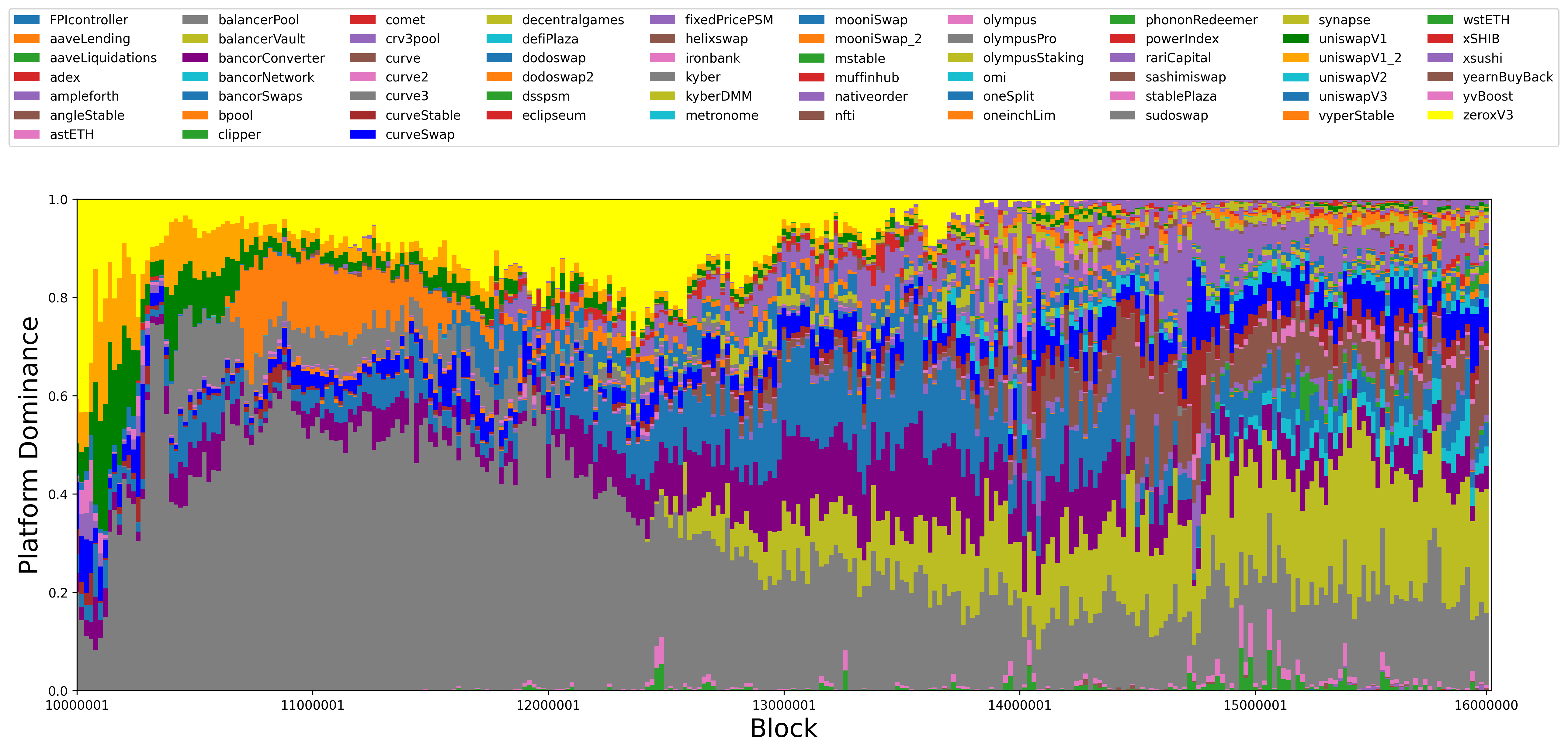}
  \caption{Dominance of each smart contract in MEV transactions with UniswapV2, UniswapV3 excluded}
  \label{fig:Contracts Ratio}
\end{figure}

Figure \ref{fig:Contracts Ratio} excludes two dominant platforms, Uniswap V2 and Uniswap V3, providing a clearer representation of the changes in emerging and declining contracts. In this figure, early-period contracts like Uniswap V1 and 0x V3 have lost their dominance, while newer contracts, such as Balancer, have experienced an increase in usage.

Figures \ref{fig:Contracts Count},\ref{fig:Contracts Ratio2}, and \ref{fig:Contracts Ratio} emphasize that the smart contracts employed for MEV transactions have increased and diversified. Alongside the growing complexity of MEV strategies, we note that detection methods relying on contracts, events, and function calls are not sustainable and require considerable maintenance.

\subsection{Sandwiches Detection Algorithm Evaluation}
Based on our labeled data, we evaluated the performance of Algorithm \ref{alg:sandwich detection algorithm} for the period covering blocks 15,500,000 to 15,650,000. The accuracy, precision, recall, and F1-score are summarized in Table \ref{table:sandwich detection algorithm performance}. The results indicate that the sandwich detection algorithm \ref{alg:sandwich detection algorithm} exhibits a high level of effectiveness in detecting sandwich attacks.

\begin{table}[h!]
\caption{Algorithm \ref{alg:sandwich detection algorithm} Performance}
\begin{tabularx}{0.45\textwidth}{cYYYY}
\toprule
 & Precision & Recall & F1 \\
\midrule
Algorithm \ref{alg:sandwich detection algorithm} & 0.9874 & 0.9893 & 0.9883 \\
Flashbots & 0.9283 & 0.9759 & 0.9515 \\
\hline
\end{tabularx}
\label{table:sandwich detection algorithm performance}
\end{table}

\subsection{ArbiNet Evaluation}

\subsubsection{Data Preparation}
Utilizing the labeled arbitrage data, we prepared a dataset for training and testing ArbiNet. For training, we utilized labeled arbitrage data from blocks 15,540,000 to 15,585,000. During this period, we identified 45,704 arbitrage transactions and labeled them as 1. To collect non-arbitrage transactions with a label of 0, we first obtained a list of transactions and filtered out those with fewer than two token transfers. This filtering process helps the model avoid classifying overly simple tasks, thereby improving its ability to distinguish between arbitrage and non-arbitrage transactions. We then randomly sampled 45,704 non-arbitrage transactions to create a balanced dataset with an equal count of labeled arbitrage transactions.

For the test data, we used transaction data from 5,000 blocks, ranging from block 15,585,001 to 15,590,000. Within this period, we found 4,200 arbitrage transactions and 90,000 non-arbitrage transactions with at least one token transfer. The test data is highly unbalanced, with arbitrage transactions accounting for less than 5\% of the total transactions. We chose to test with unbalanced data instead of a balanced set for two reasons: 1) It allows us to verify that our model performs well with actual transactions in a block, and 2) Testing with balanced data could lead to an overestimated F1 score, as the number of false positives would be reduced due to the sampling of only a small portion of non-arbitrage transactions.

\subsubsection{ArbiNet Performance}
We evaluated accuracy, precision, recall, and F1 score for the test data over 40 epochs and selected the epoch with the maximum F1 score. The evaluation results are summarized in Table \ref{table:ArbiNet Performance}.

\begin{table*}[h!]
\caption{ArbiNet Performance}
\begin{tabularx}{0.8\textwidth}{cl|YYYY|YYYY}
\toprule
&\multicolumn{1}{c}{} & \multicolumn{4}{c}{\textbf{Train}} & \multicolumn{4}{c}{\textbf{Test}} \\
&\multicolumn{1}{c}{} & Accuracy & Precision & Recall & \multicolumn{1}{c}{F1} & Accuracy & Precision & Recall & F1 \\
\midrule

\multirow{3}{*}{ArbiNet} & GCN & 0.9935 & 0.9966 & 0.9903 & 0.9934 & 0.9970 & 0.9501 & 0.9822 & \textbf{0.9659}\\
& GAT& 0.9974 &	0.9995 & 0.9953 & 0.9974 & 0.9983 & 0.9699 & 0.9915 & \textbf{0.9805} \\
& GraphSAGE & 0.9957 & 0.9996 & 0.9917 & 0.9956 &0.9987 & 0.9787 & 0.9922 & \textbf{0.9814}\\
\hline
Flashbots & & 0.7548 & 0.5461 & 0.9239 & 0.6865 & 0.9760 & 0.9446 & 0.4710 & \textbf{0.6285}\\

\hline
\end{tabularx}
\label{table:ArbiNet Performance}
\end{table*}
In the balanced train dataset, ArbiNet achieved an F1 score above 0.99 for each model using GCN, GAT, and GraphSAGE layers. However, using a balanced dataset for model evaluation may not be appropriate, as it significantly reduces false positives, leading to an inflated precision rate. When examining the test performance, the model using GCN layers performed worse compared to those employing GraphSAGE and GAT layers. This can be attributed to the fact that GCN layers only consider the connections between addresses, not their directions. The model with GraphSAGE layers achieved the best F1 score of 0.9854 among the three models.

We also compared the F1 score of Flashbots data for the same period of each training dataset and test dataset. Assuming that our labeled data serves as true labels of arbitrage transactions, Flashbots data showed a low F1 score of 0.6285 during the test period. The low F1 score stemmed from the low recall rate, which was 0.471, indicating that the model was conservative in detecting arbitrages.

The results demonstrate that ArbiNet can detect arbitrage transactions with a high F1 score and recall rate, clearly proving that ArbiNet can overcome the low recall rate issues that existing algorithms faced. Additionally, once ArbiNet is trained, it does not require knowledge of smart contract ABIs or event logs. The only input needed is token transfer graphs, easily obtained from transaction receipt data and easily-accessible ERC-20 ABI. Therefore, ArbiNet may successfully address the problems of maintenance and dependency on centralized API services.

\textbf{Exploring false negatives}
We investigated false positive transactions detected by ArbiNet, which employs a GraphSAGE layer and demonstrated the highest F1 score. Out of 4,271 detected arbitrage transactions, 4,180 were true positives, and 91 were false positives. Upon inspecting the false positive transactions, we identified arbitrage transactions that our labeled data failed to capture. One transaction\footnote{0x9a3f94c9c0e0a33055769785208d7dfc32faf77b7bae244239aa69e3297c15c7} contained an event \textit{RPLFixedSupplyBurn}, a new type of token exchange event in a Rocket Pool contract\footnote{0xD33526068D116cE69F19A9ee46F0bd304F21A51f}. Another false positive transaction\footnote{0xcf15fa9892b2ea39ac3b956e5a8af4bb8b5ab41519cc847c9a6d52e33e94c3a2} involved arbitrage using an event \textit{Withdraw} in a contract\footnote{0xF3505383b740af8C241f1CF6659619A9c38D0281}, which also represents a new type of event that triggers token exchanges.

These findings are noteworthy because ArbiNet can detect new types of arbitrages more efficiently. By identifying arbitrages based on token transfers without relying on contract information, ArbiNet exhibits the potential to rapidly recognize new threats that may adversely impact users or the blockchain network itself. 

\subsubsection{Ablation study}
In Table \ref{table:ArbiNet Performance excluding each feature group}, we conducted an ablation study by training and evaluating our model after excluding a group of node features described in Table \ref{table:Node Features}. The aim of this study is to confirm the importance of each node feature group extracted from the graph in detecting arbitrages. The feature categories of profits include $F1, F2, F3$; token categories consist of $F4, F5, F6$; address categories encompass $F7, F8, F9, F10, F11$; and transfer categories comprise $F12, F13, F14$.

\begin{table}[h!]
\caption{ArbiNet F1 excluding each feature group}
\begin{tabularx}{0.45\textwidth}{cYYY}
\toprule
\begin{tabular}[c]{@{}c@{}}Excluded\\Feature Group\end{tabular} & GCN & GAT & GraphSAGE\\

\midrule
None & 0.9659 & 0.9805 & 0.9854 \\
Profits & 0.882 & 0.9451 &  0.9527\\
Tokens & 0.9372 & 0.9791 & 0.9819 \\
Addresses & 0.8394 & 0.9124 & 0.9118 \\
Transfers & 0.9635 & 0.9652 & 0.9794\\

\hline
\end{tabularx}
\label{table:ArbiNet Performance excluding each feature group}
\end{table}
From Table \ref{table:ArbiNet Performance excluding each feature group}, we observed that excluding the feature group related to addresses and profits significantly degraded the ArbiNet performance. Excluding the transfer group resulted in a small decrease in the F1 score for all three models. When the token feature group was excluded, the GCN model performance deteriorated the most compared to the GAT and GraphSAGE models, which remained robust against excluding the profits group. Since the GCN layer does not account for the direction of edges, it is likely that the token feature group provides information related to edge directions.

These results demonstrate that each feature group serves as a crucial insight into arbitrage transactions, ultimately contributing to improved model performance when considered.

\section{Discussion and Limitations}\label{discussion}
\textbf{Labeled MEV} 
By employing the classification introduced in Section \ref{classification} and leveraging the extensive contract information provided in Appendix Table \ref{tab:token exchange events},\ref{tab:token exchange functions}, we have assembled a labeled MEV dataset for training and evaluating our model. We have taken great care to ensure the completeness of this dataset, striving to minimize both false positives and false negatives. It has shown to offer a more comprehensive representation of the overall MEV.

However, as discussed in Section \ref{Existing algorithm}, the limitations of existing algorithms also apply to our labeled MEV classification. To fully capture MEV using our classification, knowledge of smart contract ABIs and pre-defined events are required, leading to the need for constant maintenance.

\textbf{ArbiNet}
Since the trained ArbiNet does not require any knowledge of smart contracts but only token transfer graphs, it alleviates the dependency on centralized API services and the need for constant maintenance of algorithms. It also demonstrates a high precision and recall rate, as seen in Section \ref{evaluation}, which is an issue that current algorithms struggle with. We believe that future MEV transactions using new smart contracts can be easily discovered with ArbiNet, while they may remain undetected using traditional methods that rely on contract ABIs.

One limitation of ArbiNet is its lack of explainability, meaning that it is difficult to explain why a transaction is detected as arbitrage. This is due to the fact that ArbiNet is based on deep neural networks, unlike rule-based existing algorithms.

\section{Conclusion}\label{Conclusion}
In this paper, we have identified three limitations of current MEV detection algorithms and demonstrated that existing methods are not reliable for understanding the present status or analyzing the security of candidate protocols. We categorized MEV transactions and labeled them, providing a more extensive dataset than any currently available public data, contributing to the open nature of the blockchain security. 
Finally, we propose a MEV detection model using graph neural networks and a sandwich detection algorithm. Our model overcomes dependence and maintenance problem by being ABI-free, and shows high F1 score, highly improving the F1 score that previous algorithms had suffered from. Our ABI-free detection model enables accurate monitoring of blockchain networks and identification of new potential threats.

\printbibliography

\SetKwComment{Comment}{/* }{ */}
\clearpage
\appendix
\section{NFT arbitrage detection algorithm}

\begin{algorithm}
\SetAlgoLined
\KwIn{${tx}_1$}
\KwOut{$is\_arbitrage$}

$nft\_transfer\_dict = get\_ERC721\_transfers(tx)$ 
\Comment*[r]{ returns a dictionary with key : tokenID and value: transfers}

\For{$tokenId, transfers$ in $nft\_transfer\_dict.items()$}{
  \If{$len(transfers) != 2$}{
    continue;
  }
  \If{$transfers[0].to == transfers[1].from$}{
    $seller = transfers[0].from$\\
    $taker = transfers[0].to$\\
    $buyer = transfers[1].to$
  }
  \eIf{$transfers[0].from == transfers[1].to$}{
    $seller = transfers[1].from$\\
    $taker = transfers[1].to$\\
    $buyer = transfers[0].to$
  }{
    continue;
  }
  \If{$sender == seller \text{ or } sender == buyer$}{
    break;
  }
  \For{$transfer$ in $ERC-20 transfers$}{
    \If{$transfer.from == taker \text{ and } transfer.to == seller$}{
      $sell\_amount = [transfer.amount, transfer.token]$ 
    }
    \If{$transfer.from == buyer \text{ and } transfer.to == taker$}{
      $buy\_amount = [transfer.amount, transfer.token]$ 
    }
  }
  \If{$sell\_amount[1] == buy\_amount[1] \text{ and } sell\_amount[0] < buy\_amount[0]$}{
      return $True$
    }{
    return $False$
    }
}
\caption{NFT Arbitrage Detection Algorithm}
\label{alg:NFT Arbitrage detection algorithm}
\end{algorithm}

Algorithm \ref{alg:NFT Arbitrage detection algorithm} presents a method for detecting A5, the arbitrage transactions that profit from NFTs. Unlike fungible tokens, NFTs possess unique token IDs, making it considerably easier to identify A5. Moreover, incorporating ERC-721 token transfer data into ArbiNet for the sole purpose of detecting A5 would be inefficient. As such, Algorithm \ref{alg:NFT Arbitrage detection algorithm} provides a significantly more efficient approach to detect these transactions.

\section{Token Exchange events and functions}

Upon conducting an in-depth examination of MEV transactions, we have manually identified 47 events and 18 function calls associated with token exchanges. These findings are detailed in Appendix Table \ref{tab:token exchange events} and Appendix Table \ref{tab:token exchange functions}. It is important to note that the observed events and functions are subject to change as new contracts emerge and existing ones become deprecated.


\begin{table*}
  \caption{Token Exchange events}
  \label{tab:token exchange events}
  \begin{tabular}{lccl}
    \toprule
    Platform & Contract Address & Event\\
    \midrule
    Aave & 0xC6845a5C768BF8D7681249f8927877Efda425baf & LiquidationCall \\
    Angle Stable & 0x282DFfb8D0215D7eFB8d8C5fF90aED185d8850ab & BurntStablecoins \\
    astETH & 0xbd233D4ffdAA9B7d1d3E6b18CCcb8D091142893a & Burn \\
    Balancer Pool & 0x4833e8b56fC8e8A777fcc5e37CB6035c504C9478 & LOG\_SWAP \\
    Balancer Vault & 0xBA12222222228d8Ba445958a75a0704d566BF2C8	& Swap \\
    Bancor Converter & 0x4c9a2bD661D640dA3634A4988a9Bd2Bc0f18e5a9 & Conversion \\
    Bancor Network & 0x751a3E5eCd4eCD2DE4AaCE6E55ae707A9ca10255	& TokensTraded \\
    Bancor Swaps & 0x2F9EC37d6CcFFf1caB21733BdaDEdE11c823cCB0 & Conversion \\
    BPool & 0xbc338CA728a5D60Df7bc5e3AF5b6dF9DB697d942 & LOG\_SWAP \\
    Clipper & 0xE7b0CE0526fbE3969035a145C9e9691d4d9D216c & Swapped \\
    Comet & 0x75602Ddb5315CB3be0AA86c58D88E9b78737a59F & BuyCollateral \\
    Curve 3 pool & 0xbEbc44782C7dB0a1A60Cb6fe97d0b483032FF1C7 & AddLiquidity, RemoveLiquidity \\
    Curve & 0x752eBeb79963cf0732E9c0fec72a49FD1DEfAEAC & TokenExchange \\
    Curve 2	& 0x81C46fECa27B31F3ADC2b91eE4be9717d1cd3DD7 & TokenExchange \\
    Curve 3	& 0x66310ec13f36CAf5532c32B4359760592Db835Ab & Trade \\
    Curve Stable & 0x618788357D0EBd8A37e763ADab3bc575D54c2C7d & TokenExchangeUnderlying \\
    Defi Plaza & 0xE68c1d72340aEeFe5Be76eDa63AE2f4bc7514110 & Swapped \\
    Dodo Swap 1 & 0x983dfBa1c0724786598Af0E63a9a6f94aAbd24A1 & DODOSwap \\
    Dodo Swap 2 & 0xC9f93163c99695c6526b799EbcA2207Fdf7D61aD & BuyBaseToken, SellBaseToken \\
    Dsspsm & 0x89B78CfA322F6C5dE0aBcEecab66Aee45393cC5A & BuyGem, SellGem \\
    Eclipseum & 0x08e411220e47e3Fc43BFb832186ABA95108F2861 & LogBuyDai \\
    Fixed Price PSM & 0x2A188F9EB761F70ECEa083bA6c2A40145078dfc2 & Redeem \\
    FPI controller & 0x2397321b301B80A1C0911d6f9ED4b6033d43cF51	& FPIredeemed \\
    Helix Swap & 0x08dd9604D467674BFd1dE4FfF83848166628f90F & Swap \\
    Iron Bank & 0x2aC63723a576f89b628D514Ff671300801dc1702 & Mint \\
    Kyber & 0x7C66550C9c730B6fdd4C03bc2e73c5462c5F7ACC & KyberTrade \\
    Kyber DMM & 0xcE9874C42DcE7fffbE5E48B026Ff1182733266Cb & Swap \\
    Liquity & 0xA39739EF8b0231DbFA0DcdA07d7e29faAbCf4bb2 & Liquidation \\
    Metronome & 0x686e5ac50D9236A9b7406791256e47feDDB26AbA & ConvertMetToEth \\
    Mooni Swap 1 & 0x1f629794B34FFb3B29FF206Be5478A52678b47ae & Swapped \\
    Mooni Swap 2 & 0xbeabeF3fc02667D8BD3f702Ae0bB2C4edb3640cc & Swapped \\
    Mstable & 0x15B2838Cd28cc353Afbe59385db3F366D8945AEe & Swapped, Redeemed \\
    Muffin Hub & 0x6690384822afF0B65fE0C21a809F187F5c3fcdd8 & Swap \\
    Native Order & 0x1FCC3e6F76F7a96CD2b9D09F1D3C041Ca1403c57 & LimitOrderFilled , RfqOrderFilled \\
    NFTI & 0xd8EF3cACe8b4907117a45B0b125c68560532F94D & SetTokenRedeemed \\
    Olympus Pro & 0x22AE99D07584A2AE1af748De573c83f1B9Cdb4c0 & Bond \\
    Power Index & 0x26607aC599266b21d13c7aCF7942c7701a8b699c & LOG\_JOIN, LOG\_EXIT \\
    Rari Capital & 0x67Db14E73C2Dce786B5bbBfa4D010dEab4BBFCF9 & LiquidateBorrow \\
    Ren Protocol & 0xd0DA0D062d18cc70BE85Ff94afa880EcEe66EEDD & LogMint, LogBurn \\ 
    Stable Plaza & 0x3A2b8cC91aF8bf45F3Ec61E779ee1c2ba6b7E694 & Swap \\
    SudoSwap & 0xCd80C916B1194beB48aBF007D0b79a7238436D56 & SwapNFTInPair , SwapNFTOutPair \\
    Synapse & 0x1116898DdA4015eD8dDefb84b6e8Bc24528Af2d8 & TokenSwap \\
    Uniswap V1 1 & 0x97deC872013f6B5fB443861090ad931542878126 & TokenPurchase, EthPurchase \\
    Uniswap V2 & 0xd34D4916440DBa56A5719af981e646d69C9Cec0d & Swap \\
    Uniswap V3 & 0x6958648B60a778e161C85D54D9290501fDfaD0d1 & Swap \\
    Yearn Buyback & 0x6903223578806940bd3ff0C51f87aa43968424c8 & Buyback \\
    0x Protocol V3 & 0x61935CbDd02287B511119DDb11Aeb42F1593b7Ef & Fill \\ 
    
    \bottomrule
  \end{tabular}
\end{table*}

\begin{table*}
  \caption{Token Exchange functions}
  \label{tab:token exchange functions}
  \begin{tabular}{lccl}
    \toprule
    Platform & Contract Address & Function\\
    \midrule
    1inch Lim & 0x119c71D3BbAC22029622cbaEc24854d3D32D2828 & fillOrder \\
    Aave & 0xC6845a5C768BF8D7681249f8927877Efda425baf & finalizeTransfer \\
    Adex & 0xd9A4cB9dc9296e111c66dFACAb8Be034EE2E1c2C & leave\\
    Ampleforth & 0xEDB171C18cE90B633DB442f2A6F72874093b49Ef & deposit, depositFor, burnAllTo\\
    Decentral Games & 0x4b520c812E8430659FC9f12f6d0c39026C83588D & goLight, goClassic \\
    Native Order & 0x1FCC3e6F76F7a96CD2b9D09F1D3C041Ca1403c57 & fillRfqOrder \\
    Olympus & 0x184f3FAd8618a6F458C16bae63F70C426fE784B3 & bridgeBack\\
    Olympus Staking & 0xB63cac384247597756545b500253ff8E607a8020 & stake, unstake\\
    OMI & 0xBe1b2DFb095C59Da22df63DF4Bc8f92e11A2F620 & mint \\
    Phonon Redeemer & 0xfDcC959b0AA82E288E4154cB1C770C6c4e958a91 & redeem\\
    Sashimi Swap & 0xe4FE6a45f354E845F954CdDeE6084603CEDB9410 & swapExactTokensForTokens \\
    Synthetix & 0x3Ed04CEfF4c91872F19b1da35740C0Be9CA21558 & exchangeAtomically\\
    Synthetix Multi & 0x5D4C724BFe3a228Ff0E29125Ac1571FE093700a4 & burn \\
    Vyper Stable & 0x5D0F47B32fDd343BfA74cE221808e2abE4A53827 & exchange\_underlying \\
    wstETH & 0x7f39C581F595B53c5cb19bD0b3f8dA6c935E2Ca0 & wrap, unwrap \\
    xSHIB & 0xB4a81261b16b92af0B9F7C4a83f1E885132D81e4 & etner, leave\\
    xSUSHI & 0x8798249c2E607446EfB7Ad49eC89dD1865Ff4272 & enter, leave\\
    yvBoost & 0x9d409a0A012CFbA9B15F6D4B36Ac57A46966Ab9a & deposit \\
    \bottomrule
  \end{tabular}
\end{table*}

\end{document}